\pdfoutput=0
\documentclass{aa}

\usepackage {epsfig,graphicx,color}
\usepackage {txfonts }
\usepackage {natbib  }
\usepackage {float   }
\usepackage {enumerate}
\usepackage {subfig}

\begin{document}

%
%

\title {3D simulations of pillars formation around HII regions: the importance of shock curvature
}


\author{ 
P. Tremblin  \inst{1}\and 
E. Audit     \inst{1}\and 
V. Minier    \inst{1}\and   
N. Schneider \inst{1}      
       }

\institute{Laboratoire AIM Paris-Saclay (CEA/Irfu - Uni. Paris Diderot - CNRS/INSU), 
           Cente d'\'etudes de Saclay,  91191 Gif-Sur-Yvette, France    
           }

\date{
}
\offprints{P. Tremblin}
\mail{pascal.tremblin@cea.fr}

\titlerunning{Pillars Simulations}
\authorrunning{Tremblin et al.}

\abstract{}
{ Radiative feedback from massive stars is a key process to understand
how HII regions  may enhance or inhibit star  formation in pillars and
globules at the interface with  molecular clouds. We aim to contribute
to model the interactions between  ionization and gas clouds to better
understand the  processes at  work. We study  in detail the  impact of
modulations on the cloud-HII  region interface and density modulations
inside the cloud.  }
{ We  run three-dimensional hydrodynamical simulations  based on Euler
equations coupled with gravity using the HERACLES code. We implement a
method to  solve ionization/recombination  equations and we  take into
account  typical  heating  and   cooling  processes  at  work  in  the
interstellar medium and due to ionization/recombination physics.  }
{ UV radiation creates a  dense shell compressed between an ionization
front and a shock ahead.  Interface modulations produce a curved shock
that  collapses  on  itself  leading  to  stable  growing  pillar-like
structures. The narrower the  initial interface modulation, the longer
the  resulting pillar.  We  interpret pillars  resulting from  density
modulations in  terms of the  ability of these density  modulations to
curve the shock ahead the ionization front.  }
{ The shock curvature is a  key process to understand the formation of
structures  at  the  edge   of  HII  regions.  Interface  and  density
modulations  at the  edge of  the cloud  have a  direct impact  on the
morphology  of the  dense shell  during its  formation. Deeper  in the
cloud,  structures  have less  influence  due  to  the high  densities
reached by the shell during its expansion.  }


\maketitle


%
%

\section{Introduction}

Radiative feedback from massive stars might be an important process to
explain  the star  formation rates  on  galactic scales.  Its role  in
complexe structures like  giant molecular clouds is still  a matter of
debate   \citep[see][]{Dale2007,Price2010,Dale2011}.   When   the   UV
radiation from the massive  objects photoionize the surrounding gas, a
''bubble'' of hot ionized gas  expands around the star: the HII region
\citep[see][for   example]{Purcell2009}.  While   star   formation  is
inhibited inside  the bubble, the small-scale compression  at the edge
of  the HII  region, due  to its  expansion, seems  to  form elongated
structures (pillars) and globules in which the star formation activity
seems  enhanced   \citep[see][for  example]{Deharveng2010}.  Different
models  investigate  this  process.\\  First studies  of  HII  regions
\citep[e.g.][]{Stromgren1939,Elmegreen1977}  show   how  UV  radiation
leads  to  the  formation  of  an  ionization front  and  of  a  shock
ahead. The gas is compressed between them, forming a dense shell which
may lead to fragmentation,  gravitational collapse and star formation,
this is the collect and collapse scenario \citep{Elmegreen1977}. \\ An
other one  was proposed by  \citet{Bertoldi1989}, the radiation-driven
implosion  scenario. He  looks  at the  photoevaporation of  spherical
neutral clouds and finds that the ionization front drives a shock into
the cloud leading  to the compression of the  initial structure into a
compact  globule.  This  scenario  has  been studied  in  detail  with
numerical              simulations              for              years
\citep[see][]{LeflochB.1994,Williams2001,Kessel-Deynet2003}. Recently,
\citet{Bisbas2009}   and   \citet{Gritschneder2009}   looked  at   the
implosion  of  isothermal  spherical  clouds  with  smoothed  particle
hydrodynamics  (SPH)  codes.  Using  a grid  code,  \citet{Mackey2010}
produce  elongated structures  from dense  spherical clumps  using the
shadowing  effects of  these  structures.\\ In  the  last decade,  the
importance of  the initial  turbulence in the  cloud has  been studied
with       three-dimensional      simulations,       at      different
scales.  \citet{Mellema2006} present simulations  of the  formation of
the HII region with a  good agreement with observations. The interplay
between   ionization  and   magnetic  fields   has  been   studied  by
\citet{Krumholz2007} and  later by \citet{Arthur2011},  in the context
of HII region formation, finding that magnetic fields tend to suppress
the small-scale fragmentation.  On larger scales \citet{Dale2007} look
at  the impact  of ionization  feedback on  the collapse  of molecular
clouds finding a slight  enhancement of star formation with ionization
while  \citet{Dale2011}  finds almost  no  impact.  On smaller  scale,
\citet{Lora2009}  find  that the  angular  momentum  of the  resulting
clumps  is  preferentially  oriented  perpendicular  to  the  incident
radiation.   \citet{Gritschneder2010}    show   that   pillars   arise
preferentially at high turbulence  and that the line-of-sight velocity
structure of  these pillars differs from a  radiation driven implosion
scenario. \\ 

These HII regions are seen in a lot of massive molecular clouds and are the object of a large number of observation- nal studies: in Rosette nebula \citep{Schneider2010}, M16 \citep{Andersen2004}, 30 Doradus \citep{Walborn2002}, Carina nebula \citep{Smith2000}, Elephant Trunk nebula \citep{Reach2004}, Trifid nebula \citep{Lefloch2002}, M17 \citep{Jiang2002} or also the Horsehead nebula \citep{Bowler2009}. Spitzer observations provide also a wide range of HII regions studied in detail by \citet{Deharveng2010}. Pillars and globules are often seen and present density clumps which may lead to star formation. The different scenarii described above can be constrained thanks to these observations.\\

The  present  study  focuses  on  ''simple'' situations  in  order  to
highlight the key mechanisms at  work in the interaction between a HII
region and a cloud.  We  present the numerical methods needed for this
study  and then  two  different set  ups,  cloud-HII region  interface
modulation and density modulation inside the cloud and finally a study
of  the different  stage  of  evolution of  the  resulting pillars.  A
following paper  will investigate these situations in  a more complete
set up which will include turbulence inside the cloud.

%
%

\section{Numerical methods}
We consider  a molecular cloud  impacted by the  UV radiation of  a OB
cluster to study  how structures develop at the  interface between the
resulting  HII  region  and  the cloud.   Subsection  \ref{sect_hydro}
describes  the  method  used  to  simulate  gas  hydrodynamic  in  the
molecular        cloud       with       the        HERACLES       code
\citep{Gonzalez2007}. Subsection \ref{sect_io} describes the numerical
method used  to take in account  the UV radiation from  the OB cluster
and   the  resulting   ionization/recombination   reactions.   Thermal
processes from these reactions and  the heating and cooling rates used
in      the      molecular       cloud      are      described      in
subsect. \ref{thermalprocesses}.

\subsection{Hydrodynamic}\label{sect_hydro}

Our     simulations     are     performed    with     the     HERACLES
code\footnote{http://irfu.cea.fr/Projets/Site\_heracles/index.hmtl}. It
is  a grid-based code  using a  second order  Godunov scheme  to solve
Euler equations. These equations are given in Eq. \ref{eulergrav} in a
presence  of a  gravitationnal  potential $\Phi$,  constrained by  the
Poisson equation: $\triangle \Phi = 4\pi G \rho$.

\begin{eqnarray}\label{eulergrav}
\frac{\partial \rho}           {\partial t} + \nabla \cdot (\rho \textbf{V})                         & = & 0                                    \quad, \cr
\frac{\partial \rho \textbf{V}}{\partial t} + \nabla \cdot [\rho \textbf{V} \otimes \textbf{V} + pI] & = & - \rho \nabla \Phi                   \quad, \cr
\frac{\partial E}              {\partial t} + \nabla \cdot [(E + p) \textbf{V}                     ] & = & - \rho(\textbf{V} \cdot \nabla \Phi) + \Lambda(\rho,T) \quad.
\end {eqnarray}

We use an ideal gas equation of state, so that $E=p/(\gamma-1)+0.5\rho \textbf{V}^2$ with $\gamma=5/3$. $\Lambda(\rho,T)$  corresponds to  the heating  and  cooling processes
detailed  in   subsect.  \ref{thermalprocesses} leading to a polytropic gas of index around 0.7 at equilibrium (for $\rho= 500/cm^3$). All  the  other
variables have their usual meaning.

\subsection{Ionization} \label{sect_io} 
The stationnary equation governing radiative tranfer in spherical geometry is given by:
\begin{equation}\label{transfer}
\mu \partial I_\nu/\partial r + (1-\mu^2)/r\times \partial I_\nu/\partial \mu= -\kappa_\nu I_\nu + \eta_\nu
\end{equation}

$\mu$  is equal  to  $\cos\theta$  and $\theta$  is  the polar  angle,
$I_\nu(r,\mu,\nu)$  is the  spectral intensity  at a  radius r  in the
direction defined by $\mu$ at the frequency $\nu$, $\kappa_\nu(r,\nu)$
is  the absorption  coefficient  and $\eta_\nu(r,\nu)$  is the  source
term.    Radiation  is   coming  from   a  point   source   such  that
$I_\nu(r,\mu,\nu)=F_\nu(r,\nu)\times\delta_{1}(\mu)$,     in     which
$\delta_{1}(\mu)$  is a Dirac  distribution whose  peak is  at $\mu=1$
corresponding  to the direction  $\theta=0$ in  spherical coordinates.
Integrating    Eq.     \ref{transfer}     over    $\mu$    leads    to
Eq. \ref{transfer2} in which the absorption coefficient and the source
term are given by Eq. \ref{etakappa}.

\begin{equation}\label{transfer2}
\partial F_\nu/\partial r +2/r\times  F_\nu = -\kappa_\nu F_\nu + \eta_\nu
\end{equation}
\begin{eqnarray}\label{etakappa}
 \eta_\nu(r,\nu) &=& S_*(\nu)\delta_{0}(r)           \cr
\kappa_\nu(r,\nu) &=& \sigma n_{H_0} \nu^3_1/\nu^3
\end{eqnarray} 
$h\nu_1$ is  equal to 13.6eV,  $\sigma$ is the ionizing  cross section
for  photons at  frequency  $\nu_1$ and  $n_{H_0}(x)$  the density  of
neutral hydrogen  atoms. $\delta_{0}(r)$  is a Dirac  peak at  r=0 and
$S_*$ is the flux from the  central OB cluster. The radiation from the
cluster  is approximated  by an  unique source  of radius  $r_*$  at a
distance  $r_0$ of the  computational domain,  $S_*$ corresponds  to a
diluted planckian function  at the star temperature: $  S_*(\nu) = \pi
(r_*^2/r_0^2)B_\nu(T_*)$. By dividing  Eq. \ref{transfer2} with $h\nu$
to have  an equation  on the number  of photons rather  than radiative
energies and averaging $\nu$ between $\nu_1$ and $+\infty$ in order to
keep track only of the ionizing photons, we get:

\begin{equation}\label{transfer3}
1/r^2\times d (r^2 F_\gamma) /d r = -n_{H_0} \bar{\sigma}_\gamma F_\gamma   + \delta_{0}(r)\bar{S}_* 
\end{equation} 

$F_\gamma$ is now  the number of ionizing photons  per unit of surface
and time arriving in the radial direction and $\bar{\sigma}_\gamma$ is
the  average cross-section  over  the planckian  source  at $T_*$  and
$\bar{S}_*$ the rate of emission of ionizing photons by the stars. We
can get plane-parallel equations in the limit $r\rightarrow \infty$.\\

We now compute the equations for photo-chemistry. We set $X$ to be the
fraction  of ionization  $X  = n_{H^+}/n_H$  in  which $n_H=n_{H^+}  +
n_{H^0}$,  it  follows   the  dynamic  of  the  gas   as  an  advected
quantity. The variation  of protons is the number  of incoming photons
which are going to interact minus  the number of protons which will be
used for recombination. The number  of interacting photons in a volume
$dv_{cell}$ is given by the number of incoming photons threw a surface
$ds_{in}$ multiplied  by the probability  of interaction given  by the
cross-section, $dP=\bar{\sigma}_\gamma n_{H^0} dr$ leading to:
\begin{eqnarray}\label{ioni}
{d(X n_H)} &=& dn_{\gamma_{int}}-dn_{H_{rec}}\cr
dn_{H_{rec}}  &=& \beta n_{H^+}n_{e^-}dt=\beta X^2 n_{H}^2dt\cr
dn_{\gamma_{int}}  &=& F_\gamma(r)ds_{in}dtdP/dv_{cell} =\bar{\sigma}_\gamma F_\gamma n_{H}(1-X)\omega dt
\end{eqnarray} 
The   dilution  term   $\omega=(ds_{in}dr/dv_{cell})$   is  equal   to
$r_{in}^2/r_{cell}^2$  in  spherical  coordinates  and  to  1  in  the
plan-parallel limit.  $r_{in}$ is the radial position  of the entrance
surface for  incoming photons $ds_{in}$  and $r_{cell}$ is  the radial
position of  the centre of  the volume $dv_{cell}$. $\beta$  gives the
rate  of recombination  and  is equal  to $2\times  10^{-10}T^{-0.75}$
$cm^3/s$ in which $T$  is the temperature of thermodynamic equilibrium
between all the species \citep[see][]{Black1981}.

\subsection{Thermal processes}\label{thermalprocesses}

Thermal  processes are  taken in  account  by adding  the heating  and
cooling rate $\Lambda(\rho,T)$ in  the equation of energy conservation
in Eq. \ref{eulergrav}. \\

In the  ionized phase,  we consider two  processes which have  a major
impact on the thermodynamic of  the gas, the photoelectric heating due
to the massive  stars UV flux and the cooling  due to recombination of
electrons onto  protons. The energy $e_\gamma(T_*)$  given by ionizing
photons is the integrated  value of $I_\nu/h\nu (h\nu-h\nu_1)$ between
$\nu_1$  and $+\infty$  on  the incoming  spectrum  $I_{\nu}$. In  the
following we  will assume  a planckian distribution  $B_\nu(T_*)$ with
$T_* = 35  OOO °K$ which gives $e_\gamma(T_*) =  1 eV$.  Therefore the
heating rate  is given by  Eq.  \ref{heating_io}. The cooling  term is
the  loss of  the thermal  energy of  electrons used  by recombination
given by Eq. \ref{cooling_io}.
\begin{equation}\label{heating_io}
\mathcal{H}=dn_{\gamma_{int}} /dt\times e_\gamma=\omega(1-X)n_H F_\gamma \bar{\sigma}_\gamma e_\gamma
\end{equation}
\begin{equation}\label{cooling_io}
\mathcal{L}=  dn_{H_{rec}}/dt \times  k_bT/(\gamma-1)= \beta X^2 n_{H}^2 k_bT/(\gamma-1)
\end{equation}

\begin{figure}[h]

\centering
\includegraphics[width=\linewidth]{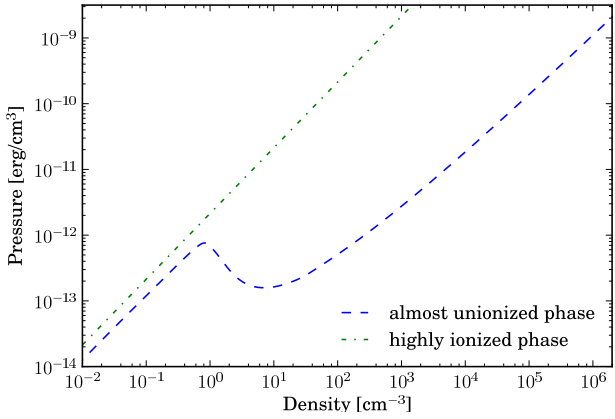}

\caption{\label{equilibrium} Thermodynamic equilibrium in the pressure-density plane for the highly and weakly ionized phases}
\end{figure}

At   the  equilibrium  between   ionization  and   recombination,  the
recombination    rate    is    equal    to   the    ionization    rate
$dn_{\gamma_{int}}=dn_{H_{rec}}$.     Therefore   when   thermodynamic
equilibrium is achieved in the  ionized phase the temperature is given
by  $(\gamma-1)e_\gamma/ k_b$  which  is equal  to  7736 Kelvin.   The
corresponding isothermal curve in  the pressure-density plane is drawn
in Fig. \ref{equilibrium}.\\

In the weakly ionized phase, we simulate the radiative heating and cooling of the interstellar medium (ISM) by adding five major processes following the work of \citet{Audit2005} and \citet{Wolfire1995,Wolfire2003}:

\begin{itemize}
\item Photoelectric heating by UV radiation with a flux equal to $G_0 /1.7$ in which $G_0$ is the Habing's flux
\item cooling by atomic fine-structure lines of C{\tiny II}
\item cooling by atomic fine-structure lines of O{\tiny I}
\item cooling by H (Ly$\alpha$)
\item cooling by electron recombination onto positively charged grains
\end{itemize}

The UV flux  used in this phase is an ambiant  low flux, additional to
the UV flux $F_\gamma$ coming from the massive stars, which is used in
our  ionization process  described in  subsect.   \ref{sect_io}.  This
heating  and cooling function  is only  valid for  the dense  cold and
weakly ionized phase. Therefore  these processes are weighted by $1-X$
and contribute  only in the weakly  ionized phase. In  this phase, the
ionization fraction  used for the  thermal processes is a  function of
the temperature and is given by \cite{Wolfire2003} (typically around 10$^{-4}$). The thermodynamic
equilibrium    in   the   pressure-density    plane   is    given   in
fig. \ref{equilibrium}.\\

The transition at the ionization front is very sharp as it can be seen on the tests in subsect. \ref{test}. Therefore the fraction of the gas in between the two phases is small and is not dynamically significant.

\subsection{1D spherical test:  HII region expansion in spherical geometry}\label{test}

\begin{figure}[h]
\centering
\includegraphics[width=\linewidth]{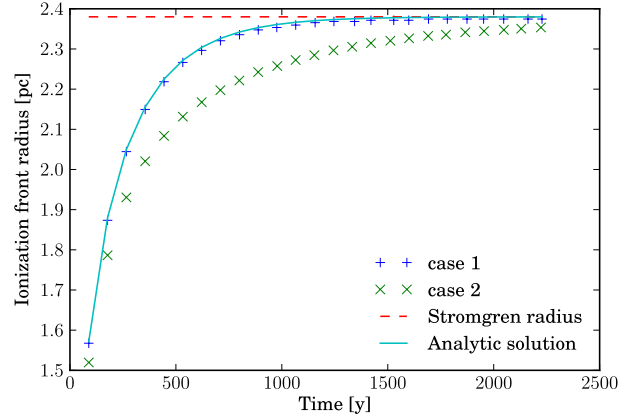}
\caption{\label{hydro_frozen}  Expansion   of  an  HII   region  in  a
homogenous    medium    (n$_H$=500    cm$^{-3}$,    T$\approx$25    K,
S$_*$=10$^{50}$  $\gamma_{ionizing}$/s).  Case  1 is  a  simulation in
which the  gas temperature is instantaneously  put at 7736  K when the
ionization fraction is bigger than 0.5.  Case 2 takes into account the
time  evolution  of temperature  given  by  Eq.  \ref{heating_io}  and
\ref{cooling_io}.  The analytic solution is computed from Eq. \ref{ri}
when  the   thermal  equilibrium  is   reached  instantaneously.   The
Str\"omgren radius is given by Eq. \ref{rs}.}
\end{figure}

\begin{figure}[h]

\centering
\includegraphics[width=\linewidth]{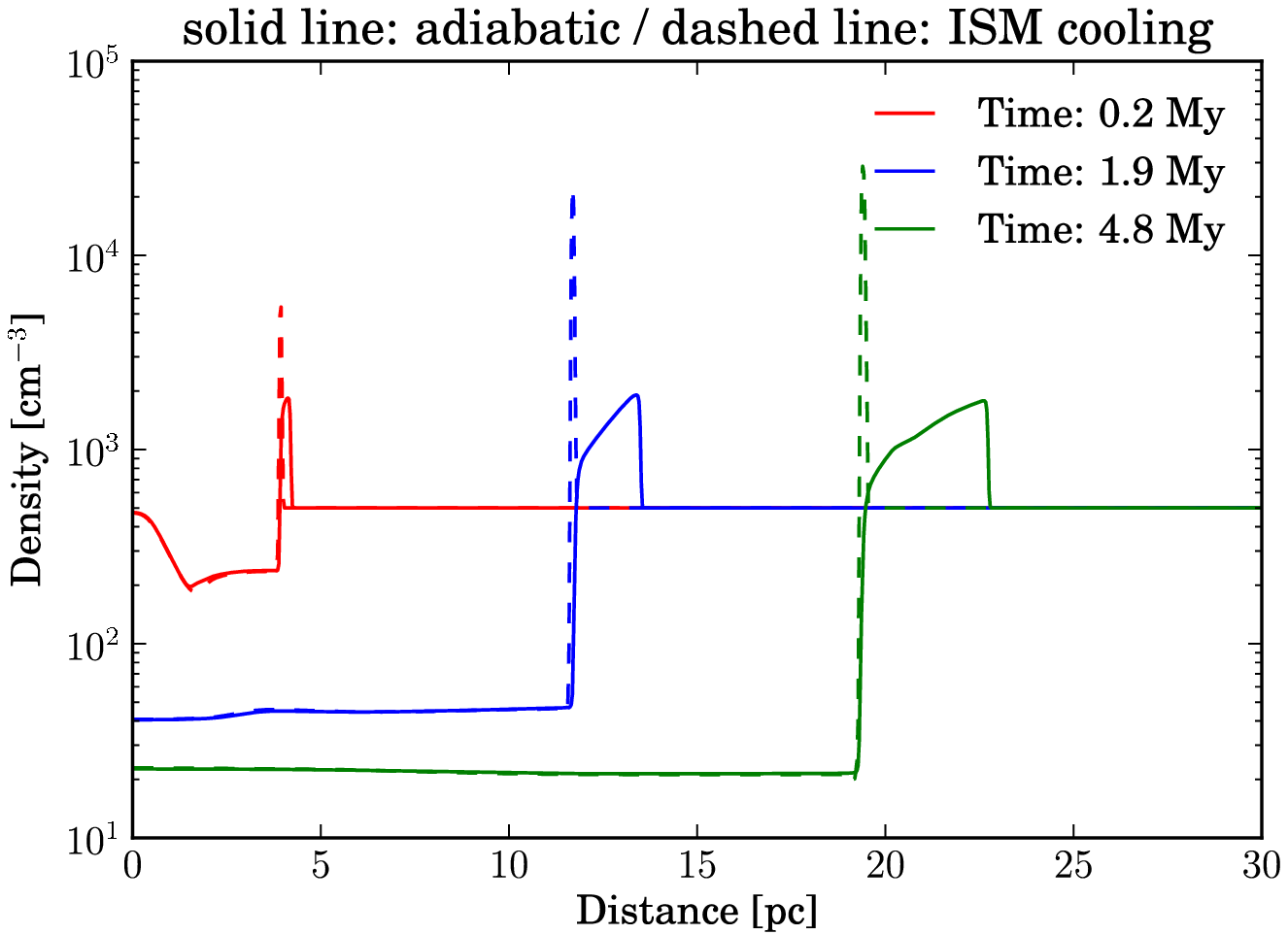}
\includegraphics[width=\linewidth]{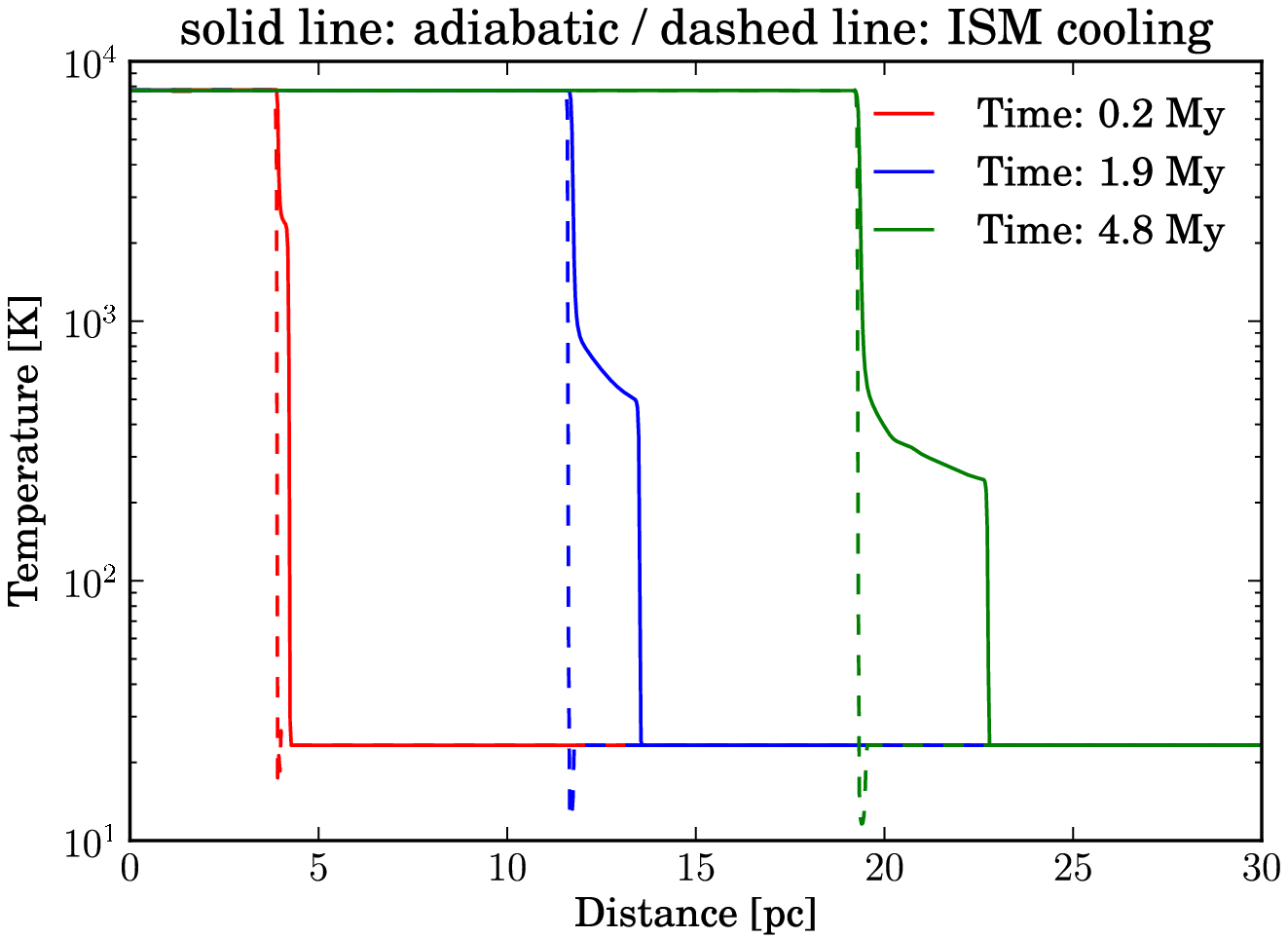}

\caption{\label{rhoT_profile}  Density (top) and  temperature (bottom)
profiles  of the expansion  of an  HII region  in a  homogenous medium
(n$_H$=500     cm$^{-3}$,      T$\approx$25     K,     S$_*$=10$^{50}$
$\gamma_{ionizing}$/s).  Solid lines correspond  to an  adiabatic cold
phase  and  dashed lines  to  a cooled  cold  phase  with the  cooling
described in Subsect. \ref{thermalprocesses}}
\end{figure}

We  first start by  testing our  numerical algorithms  on a  simple 1D
spherical test which can be compared to analytical solution.  Starting
from a  region with  n$_H$ = 500  cm$^{-3}$ and  T $\approx$ 25  K, we
switch  on  a  photon  flux  coming  from a  typically  O4  star  with
$\bar{S}_*$ =  10$^{50}$ $\gamma_{ionizing}/s$.  At  the beginning the
typical  time for  ionization/recombination is  much shorter  than the
hydrodynamical typical timescale. Therefore  there is a first phase of
development  in which  the hydrodynamic  is frozen  and the  medium is
ionized in a small sphere around the source. The radius of this sphere
is  controlled  by the  Str\"omgren  formula  given  in Eq.   \ref{rs}
\citep[see][]{Stromgren1939}.  Ionization  stops when all  photons are
used to  compensate recombinations inside  the sphere. In  our example
this radius is R$_s$ = 2.38 pc.
\begin{equation} \label{rs}
R_s = \left( \frac{3S_*}{4\pi\beta n_H} \right)^{1/3}
\end{equation}

If the thermal equilibrium between ionization and recombination is reached instantaneously, it is possible to compute an analytic expression of the time evolution of the ionization front \citep{Spitzer1978}:

\begin{equation} \label{ri}
R_i = R_s(1-e^{-\beta n_H t})^{1/3}
\end{equation}
  
The development  of the ionized  sphere before hydrodynamic  starts to
matter  is shown  in Fig.  \ref{hydro_frozen}. Two  cases  are treated
numerically, first the  thermal equilibrium is reached instantaneously
and  the gas temperature  is switched  at 7736  K when  the ionization
fraction is  bigger than  0.5. It allows  us to compare  our numerical
solution to the  analytic formula given in Eq.  \ref{ri}: the error is
less than 0.5\%.  Second we resolve numerically the  time evolution of
the temperature given by  the balance between Eq. \ref{heating_io} and
\ref{cooling_io}. The ionization front  is slowed down by the explicit
treatement  of cooling  due to  recombination. The  recombination rate
increases  when   the  temperature   decreases  leading  to   a  lower
penetration of ionizing photons.

After this  first step  the energy deposited  by the  ionizing photons
creates  a spherical  region with  a high  pressure. This  region will
therefore  start expanding  and pushing  the  surrounding interstellar
matter,  creating a  shock  front ahead  the  ionization front.   This
process creates an expanding ionized  cavity and a shell of compressed
gas.   Typical  density   and  temperature   profiles  are   given  in
Fig. \ref{rhoT_profile}. Solid lines corresponds to the adiabatic case
for  the  cold  gas and  the  dashed  lines  to the  cooling  function
described  in Subsect.   \ref{thermalprocesses}.  The  shock  front is
cooled down by the thermal processes and the shell is compressed up to
two order of  magnitude. This phenomenon is at the  origin of the idea
of  the collect  and collapse  scenario for  triggered  star formation
around HII regions proposed by \citet{Elmegreen1977}. \\

We can derive  the equations governing the density,  the speed and the
thickness  of  the  shell   in  the  approximation  of  an  isothermal
D-critical  shock   following  \citet{Elmegreen1977,Spitzer1978}.  The
density and  the pressure when  the HII region  has a radius r  can be
derived from  the equilibrium between ionization  and recombination in
the sphere and are given by:
\begin{eqnarray}\label{hii_param}
n_{II} &=& n_0 ( r_s / r)^\alpha \cr
p_{II} &=& 2 n_{II} k_b T_{II}
\end{eqnarray}
$\alpha$ is equal to 1/3 and $r_s$ to $(3 S_*/4 \pi n_0^2\beta)^{1/3}$
in spherical  geometry while, in the plan-parallel  limit, $\alpha$ is
equal to 1/2 and $r_s$ to $F_\gamma/n_0^2\beta$. The parameters of the
shell can be  computed using Eq. \ref{hii_param} and  assuming a shell
temperature $T_{shell}$ and  the corresponding sound speed $c_{shell}$
\citep[for   details,  see][]{Elmegreen1977}.   They   are  given   by
Eq. \ref{shell_param}  in which $n_{shell}$ is the  maximum density in
the shell,  $v_{shell}$ the shock  speed, and $l_{shell}$  the typical
width of the shell.

\begin{eqnarray}\label{shell_param}
n_{shell} &=& \gamma p_{II}/c_{shell}^2/(\mu m_H) \cr
v_{shell} &=& (8p_{II}/3\rho_0)^{1/2} \cr
l_{shell} &=& 4r/\gamma c_{shell}^2/v_{shell}^2
\end{eqnarray}

Assuming a  temperature in the shell of  10 kelvin and a  radius of 20
parsecs   for   the  HII   region   we   get   a  shell   density   of
3.2$\times$10$^4$cm$^{-3}$, a  speed of  3.2 km/s and  a width  of 0.5
parsec. These values  are in good agreement with  the shell parameters
we obtain in the 1D simulation in Fig. \ref{rhoT_profile}, the maximum
density   when    the   HII-region    radius   is   20    parsecs   is
2.9$\times$10$^4$cm$^{-3}$,  the shell  speed 4.4  km/s and  the shell
width 0.27 parsec.  We can see that our code can follow with very good
accuracy  the ionization  front and  the subsequent  evolution  of the
dense  shell.  We  will  therefore  now turn  our  attention  to  more
interesting 3D cases.

%
%

\section{Forming pillars}

The interstellar  medium has a  very complex structure  with important
inhomogeneities and large density fluctuations. The passage from 1D to
3D brings a lot  of new degrees of freedom. It allows  us to study how
the  shell resulting  from the  collect and  collapse scenario  may be
perturbed.  The possibility  to have  localized density  gradients has
been extensively studied with  the radiation driven implosion scenario
\citep{Mackey2010,Bisbas2009,Gritschneder2009}   and   more   recently
within a  turbulent media  \citep{Gritschneder2010}. In this  work, we
first want to  focus on more simple and  schematic situations in which
the  physical processes  at work  can be  identified and  studied more
easily. We  will therefore  study two idealized  cases. First  we will
look  at the  interaction of  the ionization  front with  a  medium of
constant density having a modulated interface. Then we will consider a
flat interface but the presence of overdense clumps in the medium.

\subsection{From interface modulations to pillars}\label{sect_con}

\begin{figure}[h]

\centering
\includegraphics[width=0.8\linewidth]{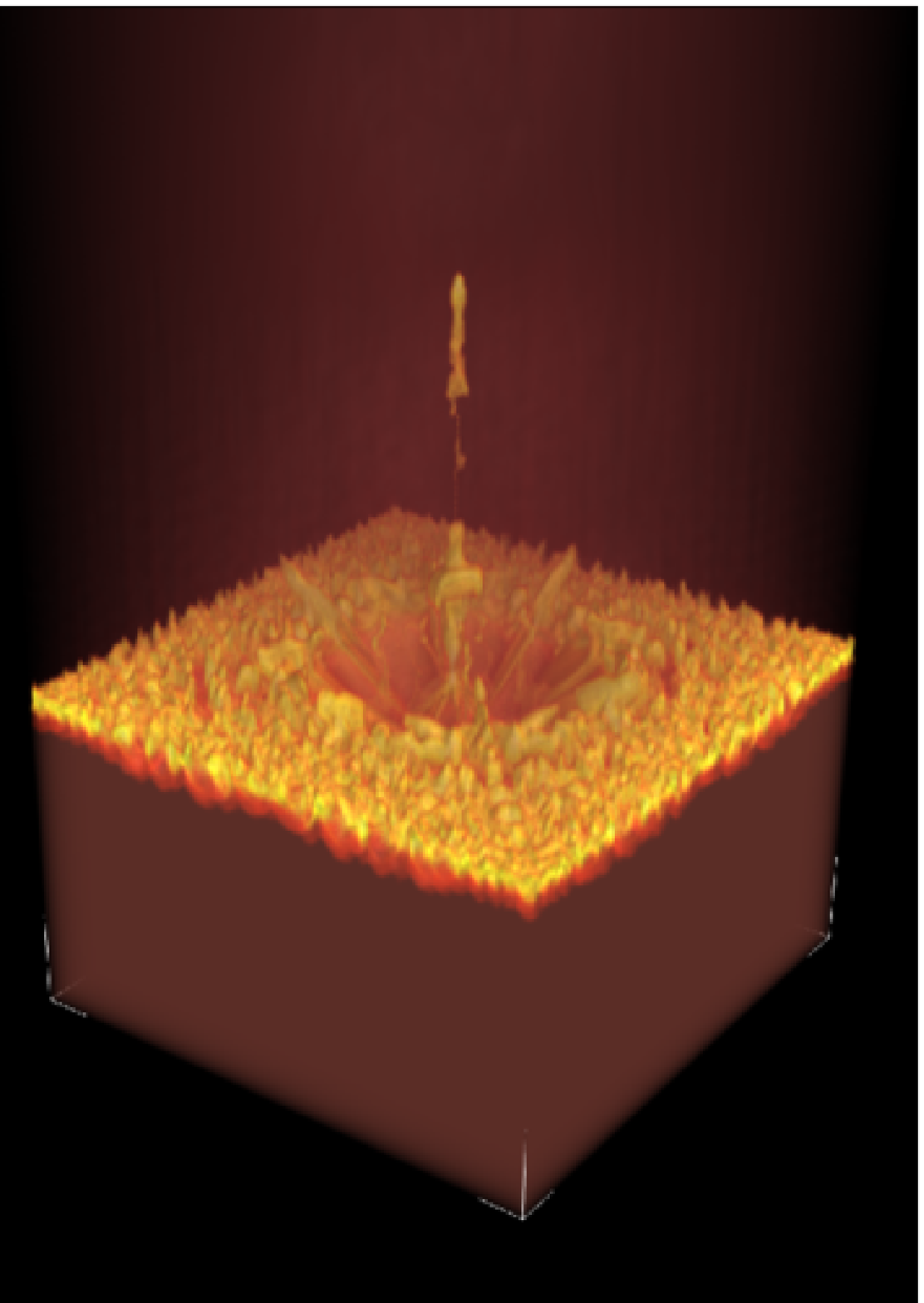}

\caption{\label{3d} 3d snapshot of density  at 500 ky of an ionization
front propagation with  a modulated interface. The UV  flux at the top
of the box is 10$^9$ photons per second per square centimetre, the box is 4x2x2 parsecs. The
dense gas  is initially at 500  H/cm$^3$ at a temperature  around 25 K
(thermal   equilibrium  from   Fig.   \ref{equilibrium}).  The   (base
width)/height  ratios  (w/h ratios  hereafter)  of  the modulation  is
w/h=0.9.}
\end{figure}

\begin{figure*}

\centering
\includegraphics[width=\linewidth]{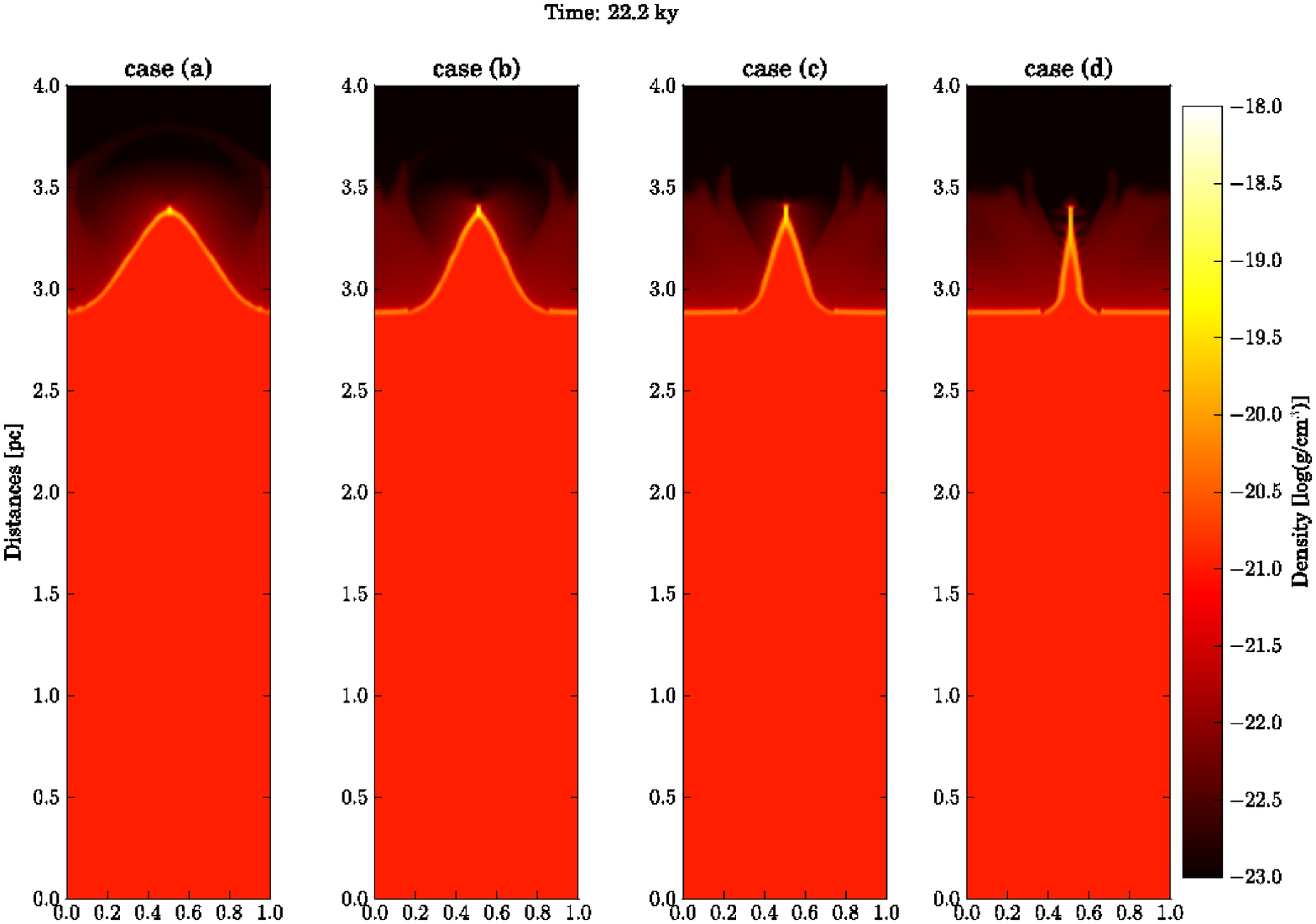}
\includegraphics[width=\linewidth]{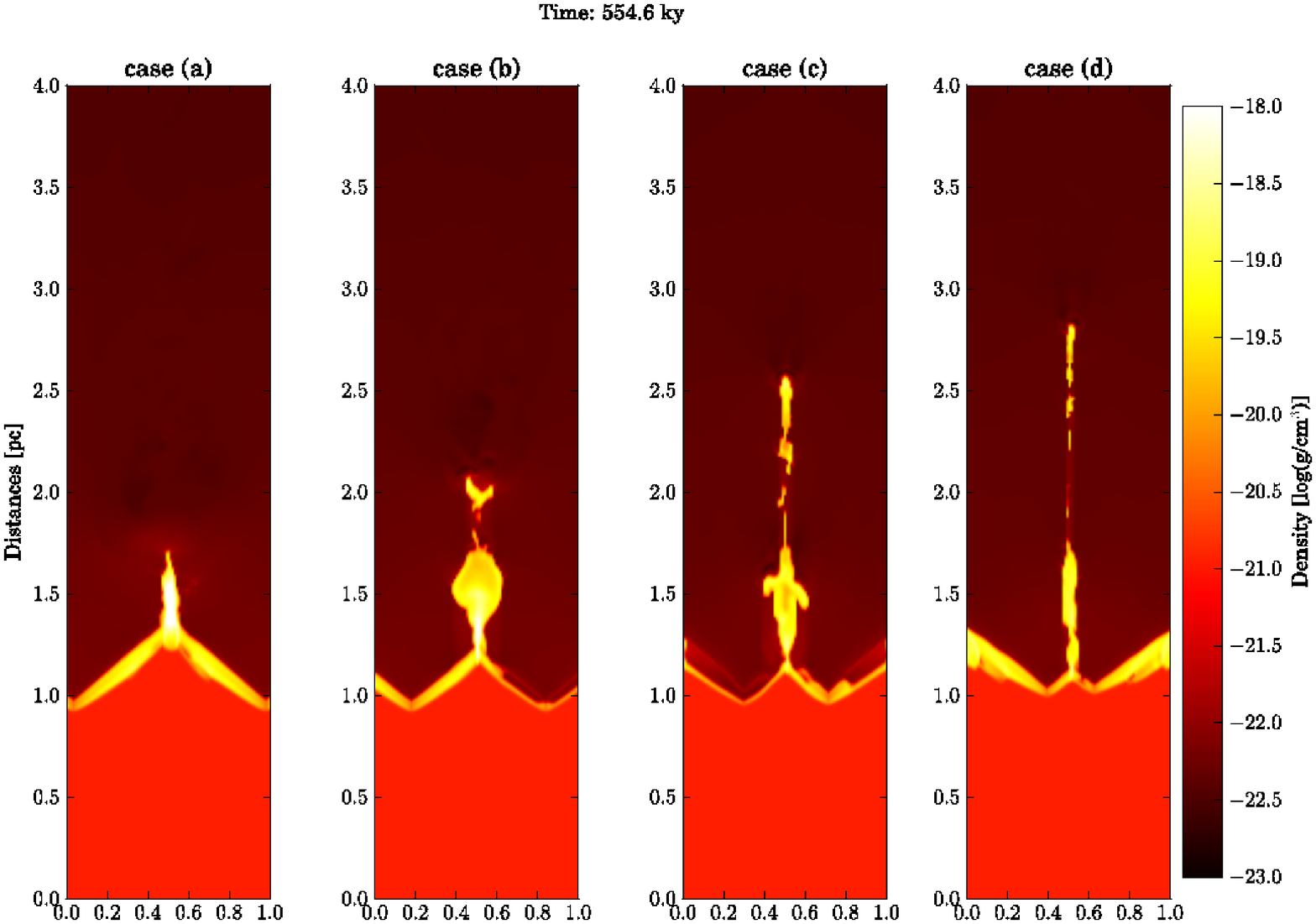}
\caption{\label{con} Density cuts of four simulations of an ionization
front propagation with  a modulated interface. The UV  flux at the top
of the  box is  10$^9$ photons  per second per square centimetre, the  box is  4x2x2 parsecs
(half of the cut is shown). The dense gas is initially at 500 H/cm$^3$
at   a   temperature   around   25   K   (thermal   equilibrium   from
Fig.  \ref{equilibrium}).  The   (base  width)/height  ratios  of  the
modulations are respectively 2 for case (a), 1.4 for case (b), 0.9 for case (c) and 0.5 for case (d). }
\end{figure*}

\begin{figure}[h]

\centering
\includegraphics[width=\linewidth]{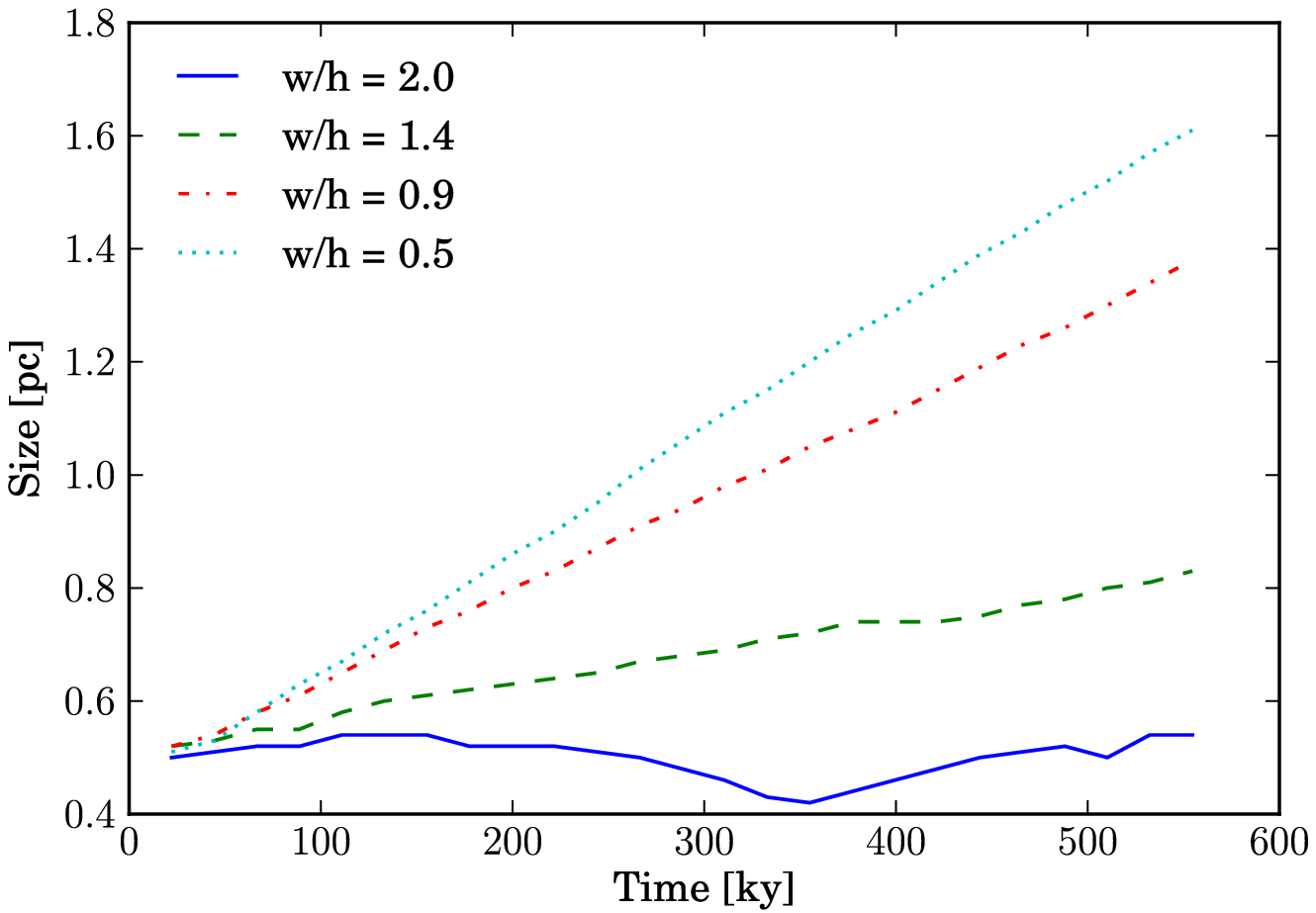}
\includegraphics[width=\linewidth]{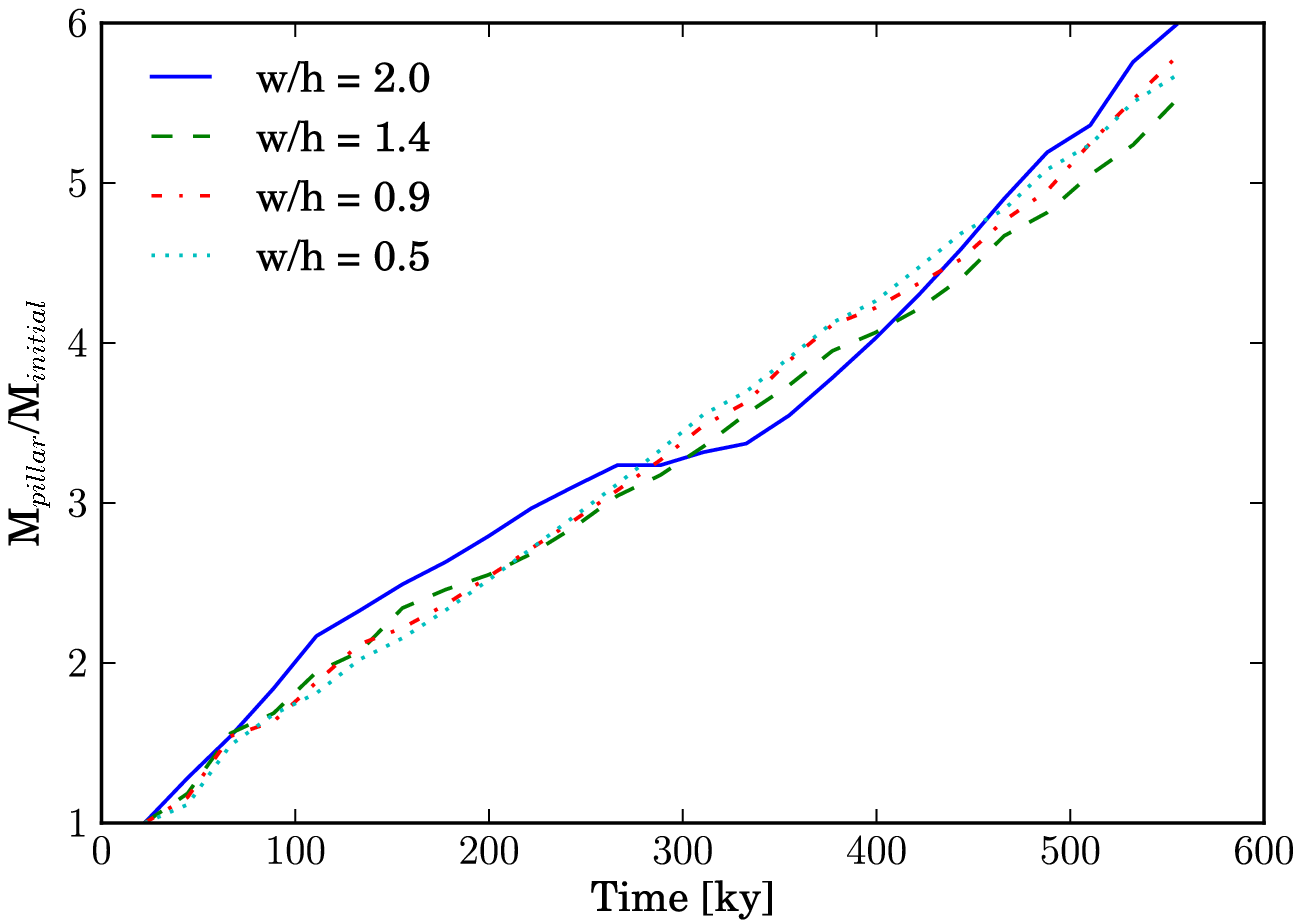}

\caption{\label{mass_size}  Monitoring of the  size (top) and  the mass (bottom) of
pillars identified  in Fig. \ref{con}.  The size is calculated  by the
difference of the average position of the ionisation front and the its
position  at  the  centre  of  the  y-z plane  (head  of  the  initial
structure). The mass  is calculated in a box  between the two vertical
positions defined to  calculated the size and with  a width defined by
the width  of the initial structure  and is normalized by  the mass in
the initial modulation.}
\end{figure}

We first  study the  interaction between the  ionization front  and an
interface modulated by an  axisymmetric sinus mode with constant heigh
(amplitude of 0.5 parsec) and   four   different   base   widths   (see   Fig.    \ref{3d}   and
Fig. \ref{con}).  The  size of the computational domain  is adapted to
the  typical observed  length scale  of pillars,  $4\times  2\times 2$
pc$^3$, at a resolution  of $400\times 200\times 200$ corresponding to
spatial  resolution  of $1\times  10^{-2}$  pc. The boundary conditions are periodic in the directions perpendicular to the ionization propagation direction, imposed where the flux is incoming and free flow at the opposite. Ionization  processes
introduced in  subsect.  \ref{sect_io} are taken  in the plan-parallel
limit.  The snapshots around 500 ky show that the narrower the initial
structures,  the   longer  the   resulting  pillars.   With   a  (base
width)/heigh ratio  (w/h ratio  hereafter) of 0.5  we get  a structure
whose  size was  almost  multiplied by  3.5  in 500  ky.  Besides  the
initial structures have less and  less mass with decreasing widths and
still  manage  to  form  longer  pillars  as can  be  seen  on  figure
(\ref{con}).  High  densities or high  initial mass are not  needed to
form  structures  which  are  going  to resist  the  ionization.   The
important  factor here  is  how  matter is  distributed  in space  and
interact  with the  propagating  shock.  Very  little and  low-density
material well  distributed in space  can result in a  long pillar-like
structure. We will call these structures pillars hereafter, however a detailed comparison with observations is needed to see whether these structures emerging from idealized scenarii are a good approximation of the reality or not.\\

To study  the properties  of the pillars,  we monitor their  sizes and
masses in Fig. \ref{mass_size}. The size gain identified previously is
a continuous process in time at an almost constant speed which depends
very strongly on the width/heigh ratio. Narrow initial modulation have
a very fast  growth while larger one grow slowly, if  at all. The mass
increase relative to  the initial mass of the  structure is almost the
same in  all simulations and does  not seems to depend  on the initial
w/h ratio. It reaches a factor of 5-6 in 500 ky. \\

\begin{figure}[h]

\centering
\includegraphics[width=\linewidth]{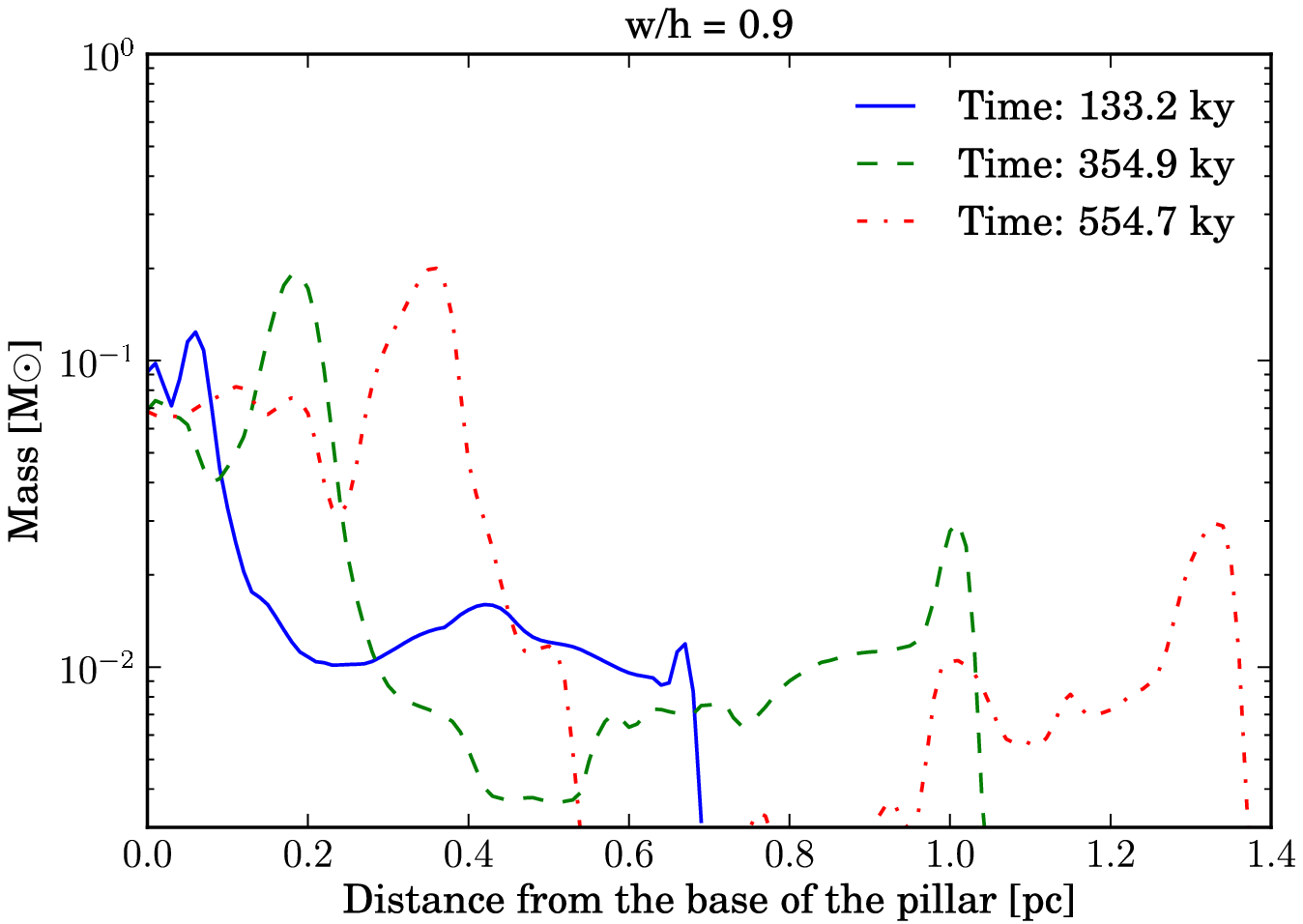}
\includegraphics[width=\linewidth]{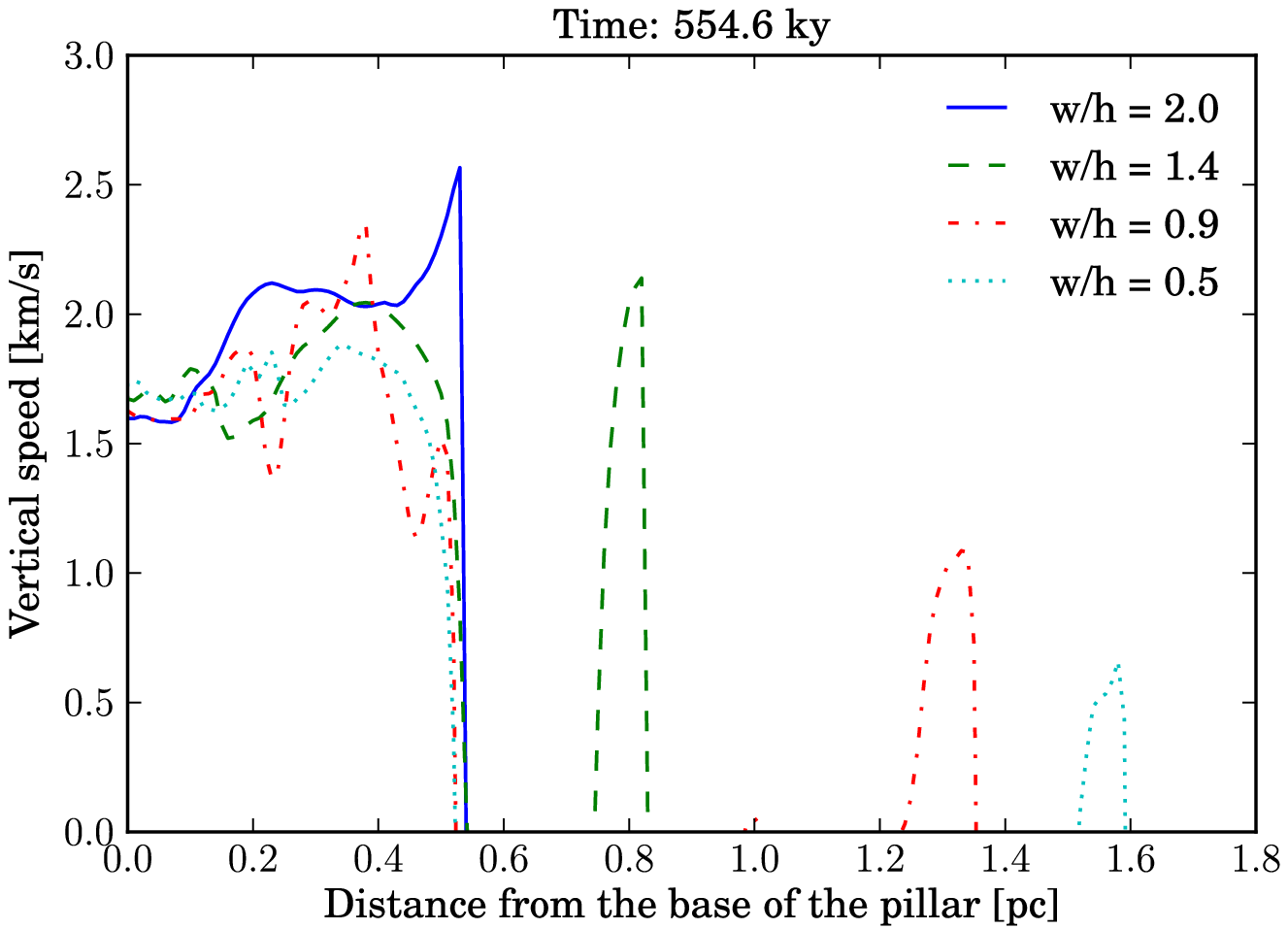}

\caption{\label{mass_profile} Mass  profiles (top) and vertical velocity profiles (bottom) at three  different times
of the  pillar in  the w/h =  0.9 simulation. 
These profiles are made within the boxes defined in Fig. \ref{mass_size}}.
\end{figure}

Vertical    profiles   of    the    mass   in    the   pillars    (see
Fig.  \ref{mass_profile})  show  that  the  mass  gain  identified  in
Fig. \ref{mass_size} is accumulated at  the base. The mass of the head
of the  pillar slightly increases  and then remains stable  during the
simulation while the central part  connecting the head and the base is
stretched and its  mass decreases. The profiles of  the vertical speed
show that the  bases of the pillars have a  roughly constant speed and
the  size variations  are  due  to vertical  speed  difference of  the
heads.  The  differences between  the  sizes  of  the pillars  can  be
directly  deduced  from  the  velocity  differences of  the  heads:  a
velocity  difference  of  0.5  km/s  during 500  ky  gives  a  spatial
difference  of  0.25  pc  which  is  approximatively  the  differences
observed between the simulations with a width/heigh ratio of 2 and the
one with a ratio of 1.4.\\

\begin{figure*}

\centering
\includegraphics[width=\linewidth]{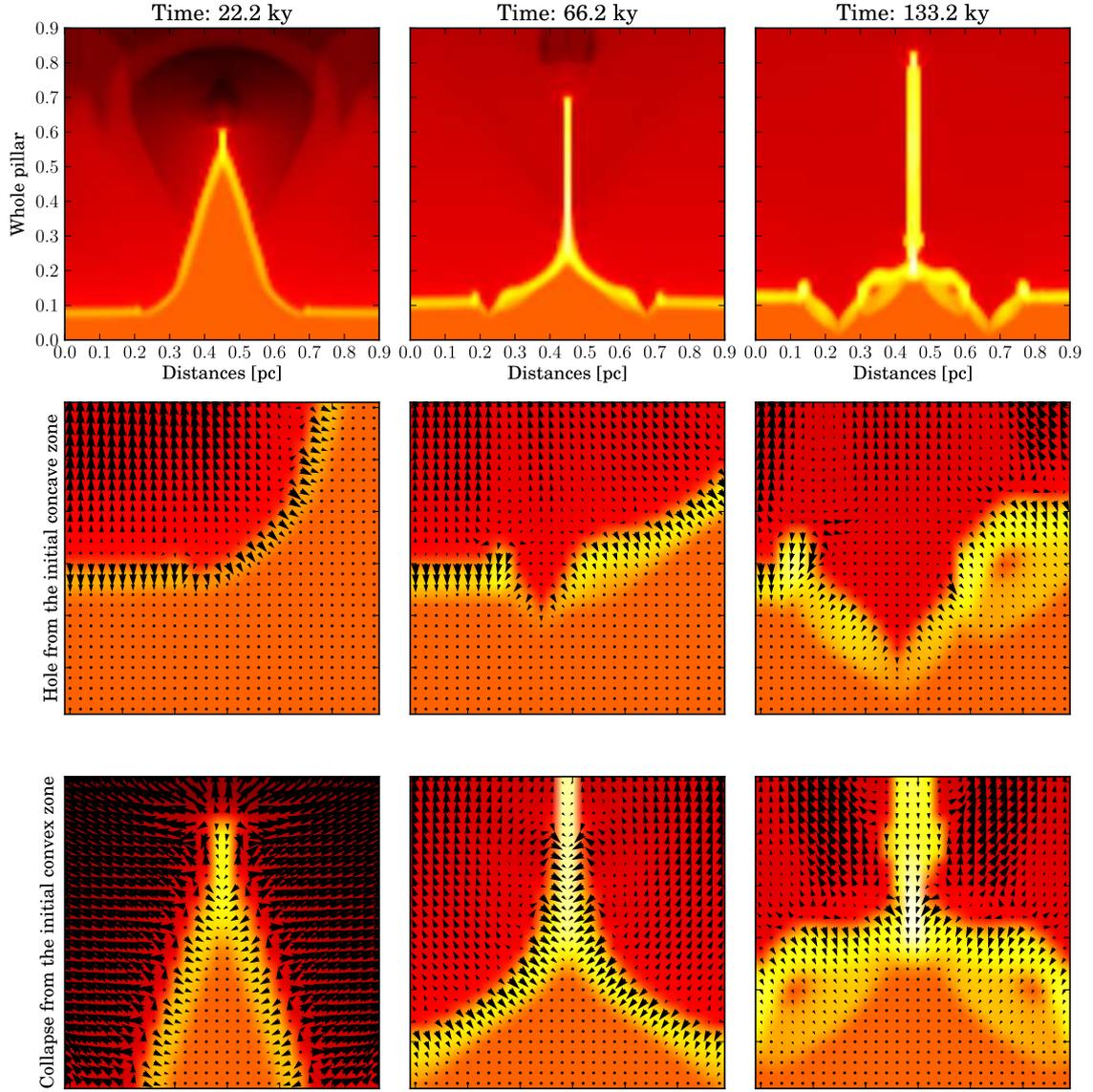}

\caption{\label{curve} Zoom  on the  ionization of convex  and concave
zone  in the  w/h  = 0.9  simulation  between 22  and  132 ky.  Arrows
represent the velocity field projected in the x-y plane.}
\end{figure*}

A closer look at the simulations (see Fig. \ref{curve}) shows that the
motions  perpendicular to  the propagation  direction are  shaping the
structures. These  motions are triggered  by the initial  curvature of
the structure,  two cases are distinguished, an  initially convex zone
and a  initially concave  zone. A  shock front on  a convex  zone will
trigger its  collapse, it is the  phenomenon which takes  place in the
radiation  driven implosion  scenario  and  here at  the  head of  the
initial  structure. The curvature  of the  initial structure  leads to
lateral  convergent shocks  that  collide on  themselves. However  the
ionization of a concave zone is quite different, it will dig a hole in
the medium, and the gas is pushed away lateraly. The velocity fields in
Fig.  \ref{curve} illustrate  these phenomena,  on a  initially convex
zone  the velocities point  inwards and  lead to  the collapse  of the
central structure  whereas on the  initially concave zone at  the base
they point outwards  and dig a hole in the  media. These phenomena are
at the origin  of the size and mass increase.  Indeed, the collapse of
the structure takes  more time if the structure  is wide, therefore it
will be accelerated down longer, explaining the difference of vertical
speed     of    the     heads     of    the     pillars    seen     in
Fig. \ref{mass_profile}. Besides, the holes on each side of the pillar
(see  Fig. \ref{curve})  gather the  gas at  the base  of  the pillar,
explaining the mass increase in Fig. \ref{mass_profile}.\\

\begin{figure}[h]

\centering
\includegraphics[width=\linewidth]{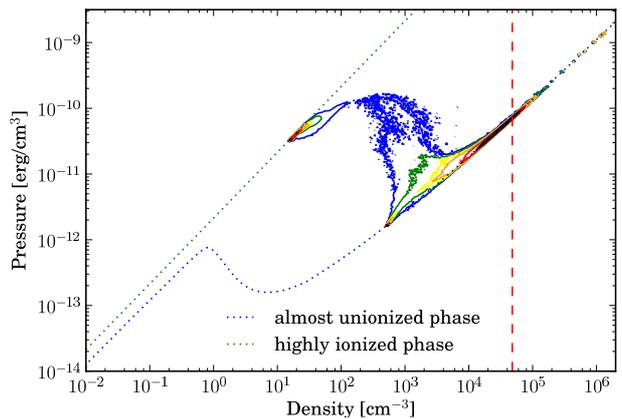}

\caption{\label{histo_cs03} 2D plot of the mass fraction at a given density and pressure for
the w/h =  0.9 simulation at the end of the  simulation (554 ky). Blue
to black contours are increasing mass fraction contours (blue: 1E-6, green: 1E-5, yellow: 5E-5, orange: 1E-4, red: 5E-4, black: 2E-3). The dashed-red line is
the  maximum  density achieved  in  a  plan-parallel  1D simulation  :
4.8$\times$10$^4$ cm$^3$. The total mass at a density above this limit
is around 10 solar masses}
\end{figure}

Besides the  mass increase,  there is an  important density  gain. The
mass histogram in the  pressure-density plane in Fig. \ref{histo_cs03}
shows  that the  compression  due  to the  heating  from UV  radiation
increases  the pressure  of one  order of  magnitude from  the initial
state  which is  at  a density  of 500  cm$^3$  and at  a pressure  of
10$^{-12}$ erg/cm$^3$.  The  gas is then distributed in  two phases at
equilibrium,  hot-ionized  on   the  green-dashed  straight  line  and
cold-dense-unionized on  the blue dashed curve. The  gas in transition
between them  is seen with the  blue contour.  A 1D  simulation of the
collect  and   collapse  process  can  increase  the   density  up  to
5$\times$10$^4$ cm$^3$, this  limit is drawn with the  red dashed line
in Fig. \ref{histo_cs03}.  The mass at  the base of the pillar, due to
the holes from the initial concave  parts and from the collapse of the
lateral  shocks, leads  to  a  density enhancement  of  1-2 orders  of
magnitude.   Although  the structures  in  our  simulations are  still
slightly under their  Jeans lengths, this process has  a better chance
to trigger star  formation at the base of the  pillars than the simple
collect and collapse scenario.

%
%

\subsection{From density modulations to pillars}

\begin{figure*}

\centering
\includegraphics[width=\linewidth]{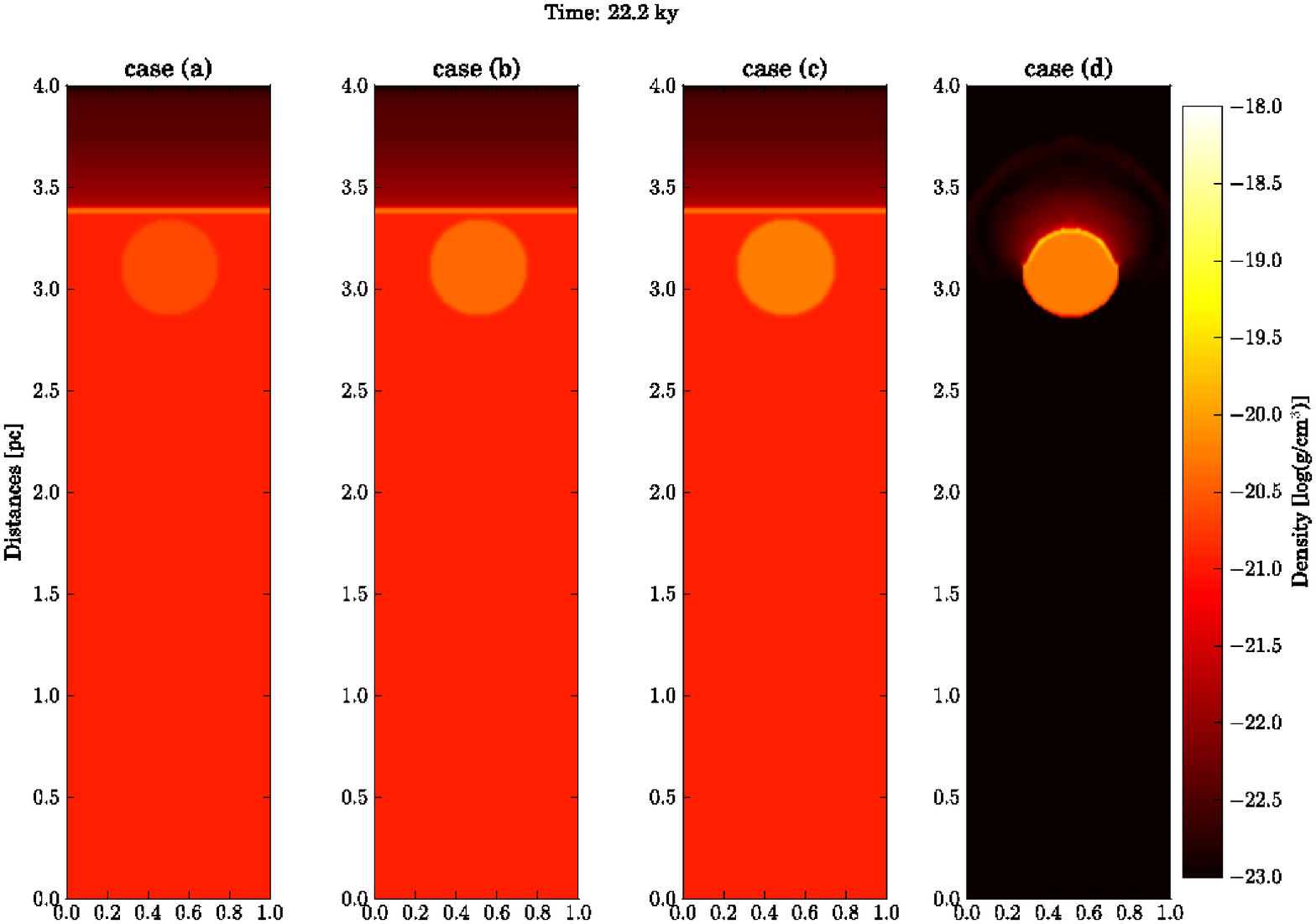}
\includegraphics[width=\linewidth]{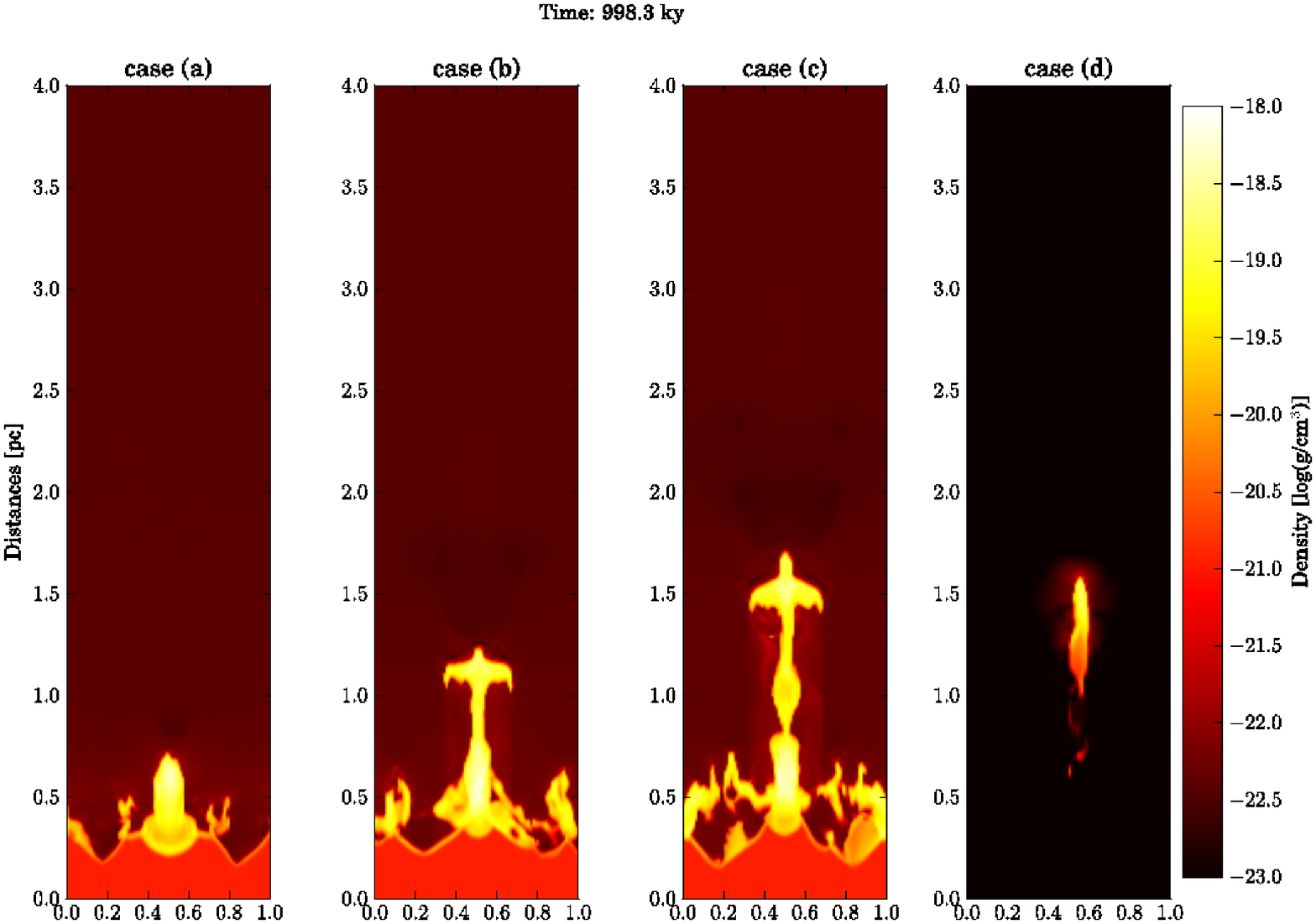}

\caption{\label{clumps}  Density  cuts   of  four  simulations  of  an
ionization front propagation on clumps (0.5 parsec in diameter). The  UV flux at the top of the
box is  10$^9$ photons per second per square centimetre,  the box is 4x2x2  parsecs (half of
the cut is shown, only the  central parsec of the simulation box). The
last simulation (case(d)) is an isolated  spherical clump of  constant density
1000 cm$^3$,  the three others are  clumps in a  homogeneous medium of
density 500 cm$^3$.  The clumps have densities of 1000 for case (a), 1750 for case (b) and 2500
cm$^3$ for case (d).}
\end{figure*}

Clouds    have     irregular    shapes    but     they    have    also
unhomogeneities. Therefore, after this study of interface modulations,
we consider density modulations that we call clumps hereafter (with no
reference to  an observational definition). We define  them as regions
of constant overdensity. Figure \ref{clumps} shows the different cases
we studied.   The three first cases  are clumps in  a homogeneous cold
cloud, with  a density contrast of  respectively 2, 3.5 and  5 and the
last one  corresponds to the radiation driven  implosion scenario, the
clump is  ``isolated'' in a hot  low-density medium so  that the shock
forms directly on the structure, the  density in the clump is the same
as the one with  a density contrast of 5. In all  cases the clumps are
at pressure equilibrium with the surrounding matter.

In the isolated case, the shock  forms at the surface of the clump and
the shell is therefore initially curved by the shape of the clump with
a w/h ratio  of 2. This roughly corresponds to  the wider case studied
in sect. \ref{sect_con}.  Since the  shock is curved, it will collapse
on itself and  form a pillar-like structure, this  case was studied in
detail  by   \cite{Mackey2010,  Gritschneder2009}.  However,   with  a
widht/heigh ratio  of 2  the structure is  quite small and  look quite
similar to the w/h=2 case  of the previous section.  However, since it
is isolated, the  accumulation of mass process at  the base identified
in the previous section can not take place.

In  the three  other  cases, the  shell  is flat  when  it is  formed,
therefore the outcomes of the  simulations are not so clear.  When the
clump is not dense enough (first case nH$_{clump}$/nH$_{cloud}$=2), it
will  not curve  the shell  enough to  have lateral  colliding shocks.
Therefore in this case, there is  no head in the final structure which
is very small (i.e comparable to  the initial clump size).  In the two
other  simulations (nH$_{clump}$/nH$_{cloud}$=3.5  and 5),  the clumps
curve the  shell enough  to trigger the  collision of  vertical shocks
thereby forming  an elongated  pillar. This is  very close to  what we
observed previously on interface modulations. Besides, we can identify
on  the final  snapshots a  base for  the pillar-structures  formed by
lateral holes in  the cloud and the associated  accretion mass process
discussed above.\\

The importance of the curvature  effect can be emphasized by comparing
the isolated  case and the nH$_{clump}$/nH$_{cloud}$=5  case.  In both
cases the density of the clump  is the same, however when the shell is
formed flat on  a homogeneous medium, the dense  clump will resist the
shock to  form an elongated modulation  on the shock  surface.  In the
isolated  case, the  shock  is formed  instantaneously  curved on  the
structure with a w/h ratio of 2. Furthermore, contrary to the isolated
case,  the  structures will  grow  in mass  and  size  because of  the
connection  with   the  cloud.   These  effects  result   in  a  final
0.5-parsec-long  pillar  for  the  isolated  case  whereas  the  final
structure  in  the  nH$_{clump}$/nH$_{cloud}$=5  case  is  1.5  parsec
long.  The  driving  process  to  form  a pillar  is  how  the  matter
distribution is  able to first  curve the shell  and then to  feed the
base of the pillar by the hole mechanism identified in sect. \ref{sect_con} (the holes are clearly visible in the final snapshots in Fig. \ref{clumps}).\\

\begin{figure}[h]

\centering
\includegraphics[width=\linewidth]{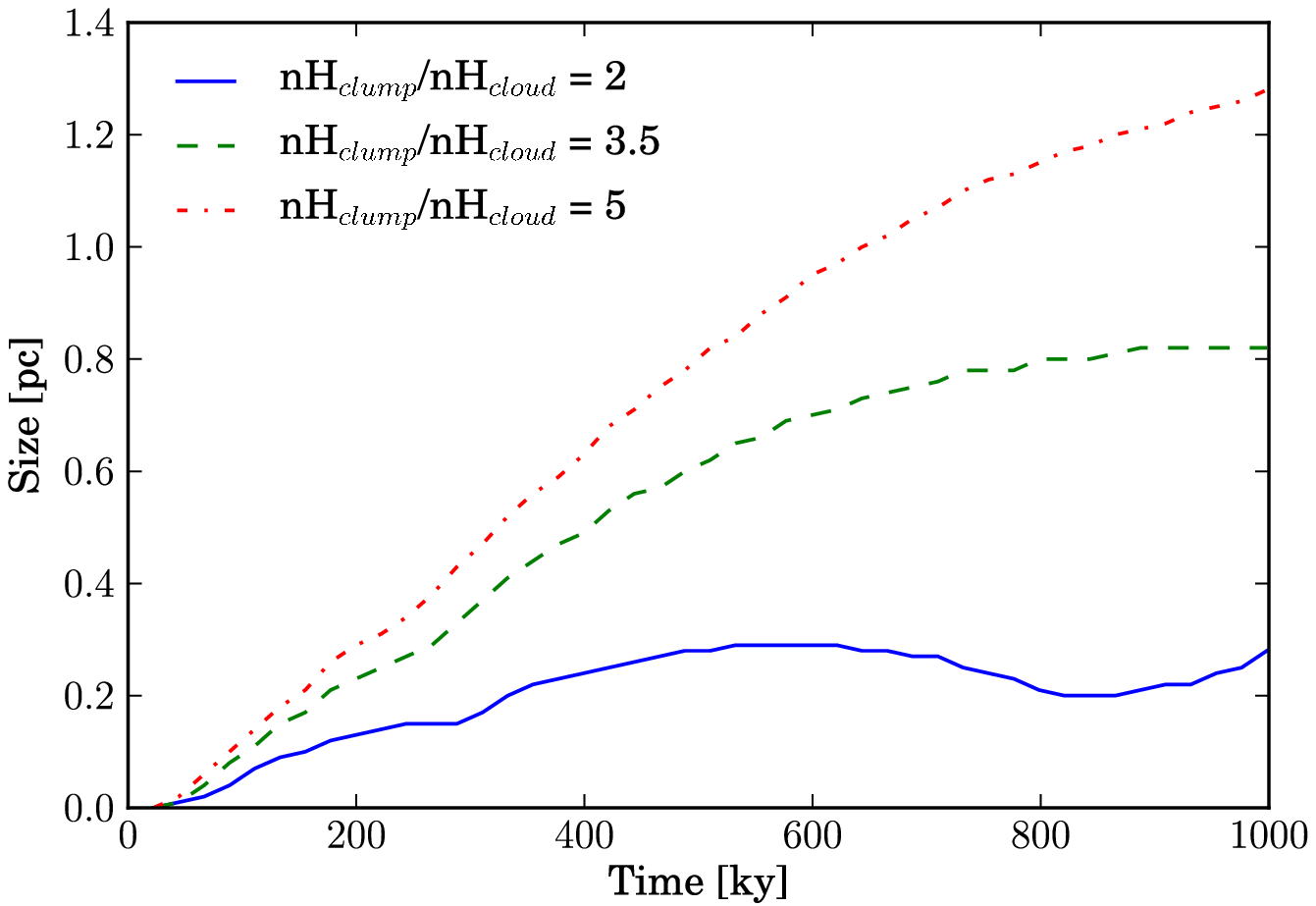}
\includegraphics[width=\linewidth]{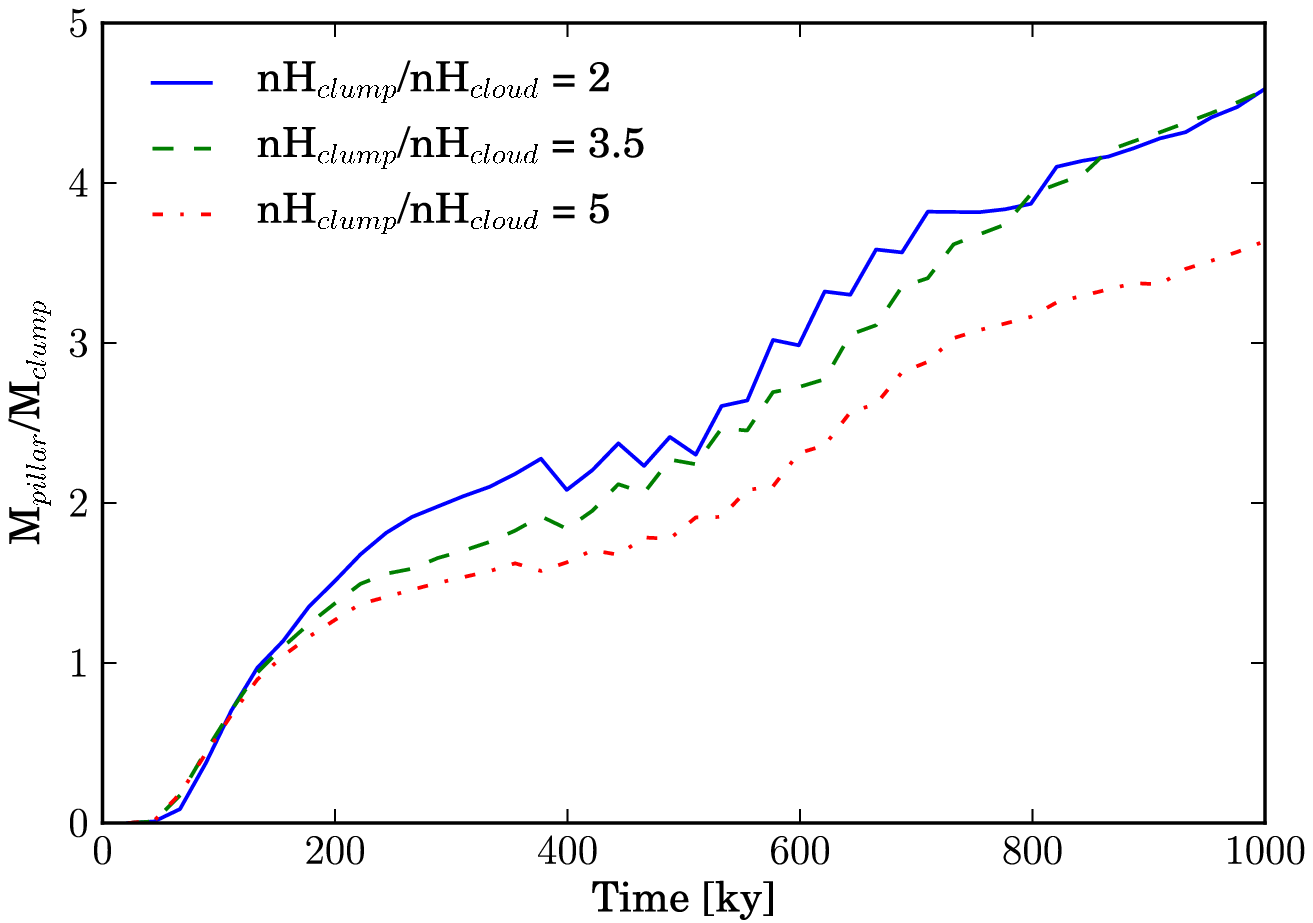}

\caption{\label{mass_size_cmp} Monitoring of the  size (top) and the mass (bottom) of
pillars identified in Fig. \ref{clumps}. The size is calculated by the
difference of the average position of the ionisation front and the its
position  at  the  centre  of  the  y-z plane  (head  of  the  initial
structure). The mass  is calculated in a box  between the two vertical
positions defined to  calculated the size and with  a width defined by
the width  of the initial structure  and is normalized by  the mass of
the initial clump.}
\end{figure}

The  size  evolution  and  the  mass  evolution  of  the  pillars  are
comparable to  the modulated-interface case. However  the evolution is
not linear  due to  the initial conditions.  Indeed, there is  a first
phase in which the ionization  front is curved and stretched verticaly
around  the  clump  when  its   propagation  is  slowed  down  by  the
overdensity. At this  point (around 200 ky) the  physical situation is
comparable to  the interface-modulated  case, the ionization  front is
curved  around a  ''hill''. It  also explains  why we  choose to  do a
longer  simulation for  the density-modulated  case, it  allows  us to
compare  both  cases  at  a  ''physically equivalent''  state  at  the
end.  When the  lateral shocks  collide  around the  hill, the  pillar
captures a  bit more than the mass  of the initial clump  and its mass
increase  slows down.  This  phase occurs  around  300 ky  and can  be
clearly  identified   in  Fig.  \ref{mass_size_cmp}.   Then  the  mass
increases  due  to the  accumulation  of matter  at  the  base of  the
pillar.   The  process is  less   linear  than   in  the
interface-modulated case. \\

\begin{figure}

\centering
\includegraphics[trim=1.8cm 1.6cm 1.5cm 1.6cm,clip,width=0.49\linewidth]{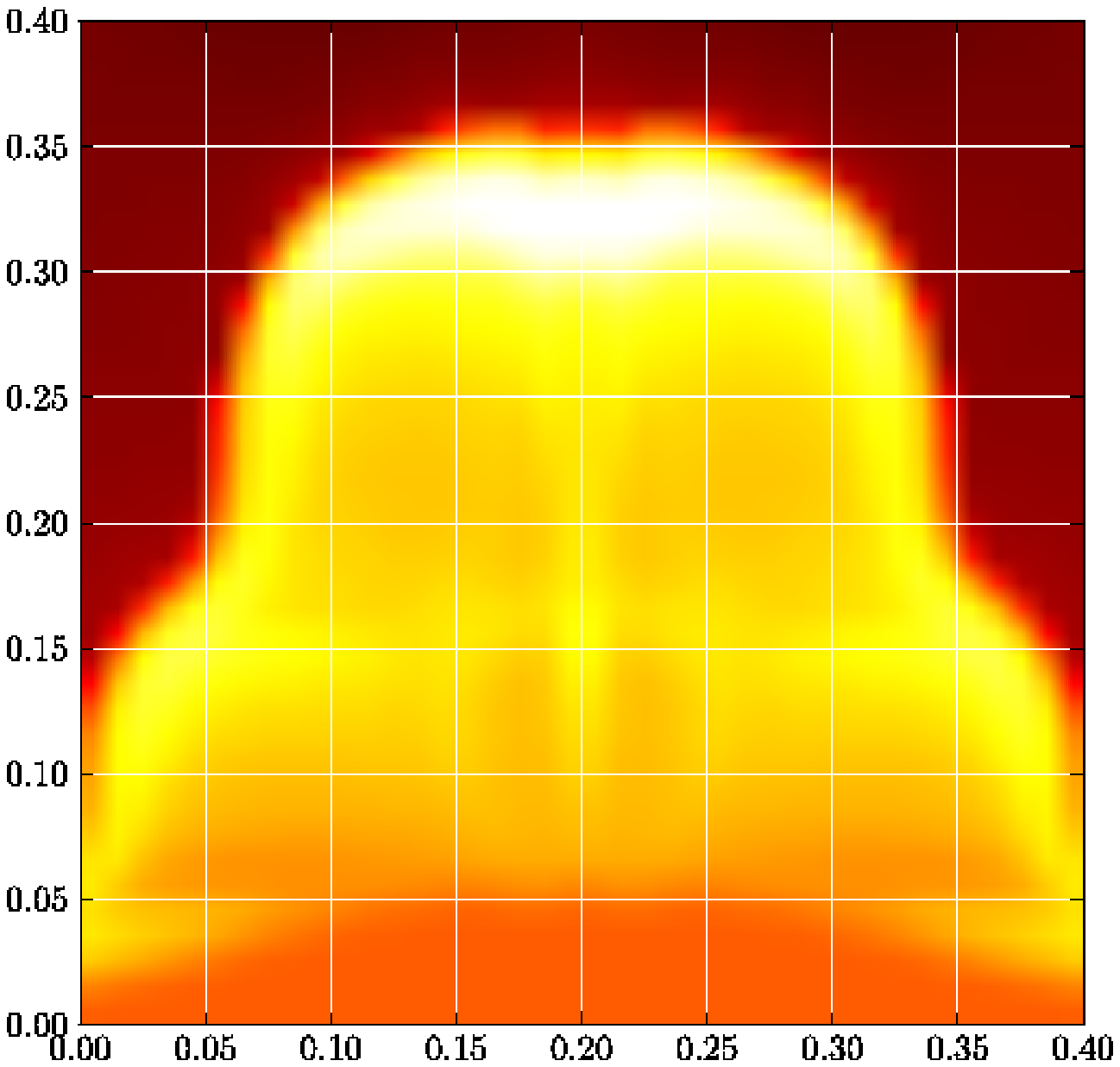}
\includegraphics[trim=1.8cm 1.5cm 1.5cm 1.5cm,clip,width=0.49\linewidth]{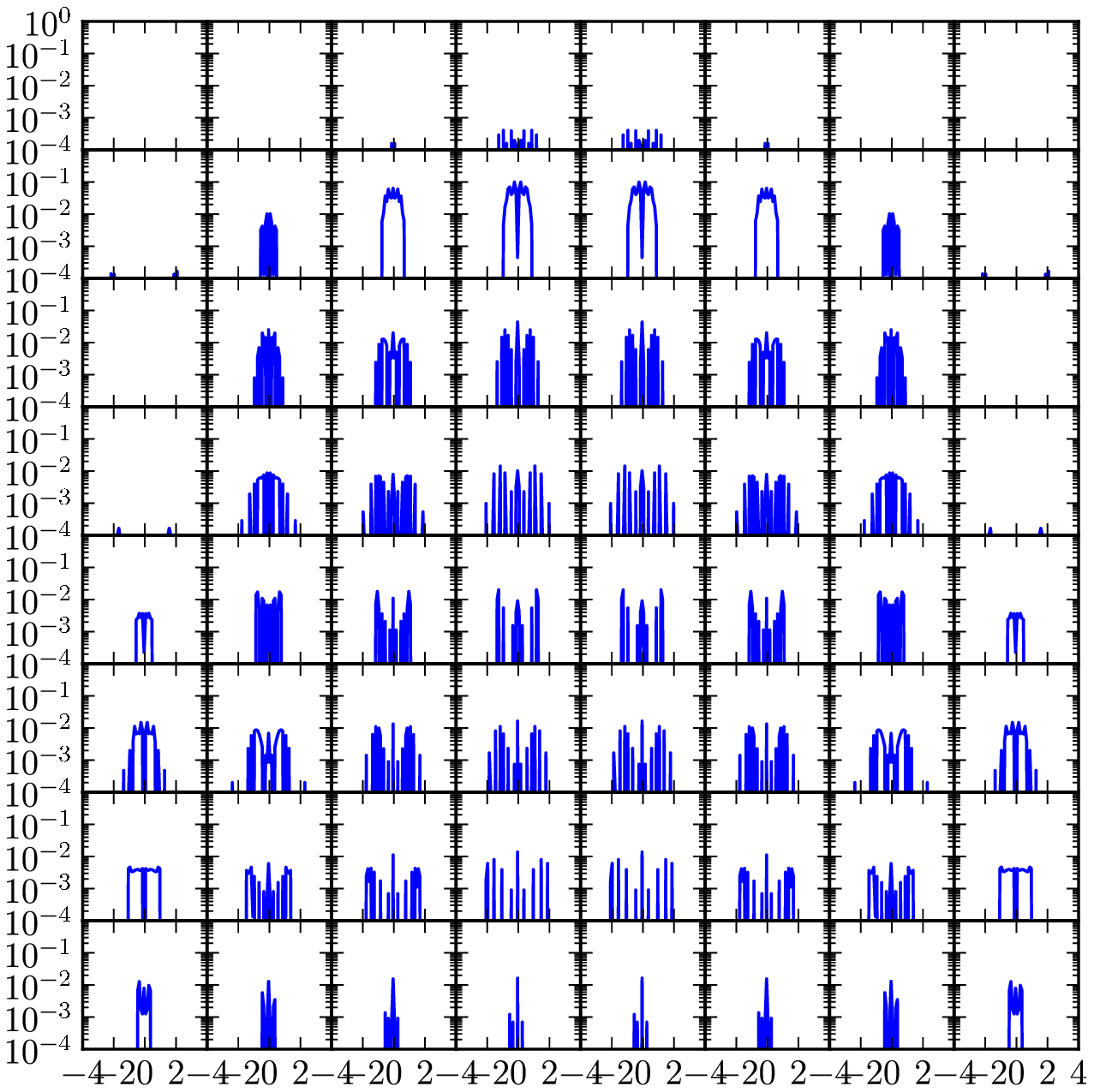}

\caption{\label{spectrum_start} nH$_{clump}$/nH$_{cloud}$=5 simulation
at t=222 ky.  Left: column density on a  0.4x0.4x0.4 pc$^3$ box around
the  pillar  structure in  a  face-on  geometry. Right:  mass-weighted
histogram of  the line of sight  velocity in the same  box (similar to
optically-thin observational line spectra). Each spectrum is made on a
square  of 0.05x0.05  pc$^2$  drawn  on the  column  density map.  The
spectra are drawn  between -4 and 4 km/s in  80 bins (horizontal axis)
and the mass between 10$^{-4}$ and  1 solar mass (vertical axis in log
scale). The  lateral shocks  can be clearly  identified in  the double
wings spectra leading to a very broad line width.}
\end{figure}

\section{Observational signature}

\begin{figure}

\centering
\includegraphics[trim=1.8cm 3.3cm 1.5cm 3.3cm,clip,width=0.49\linewidth]{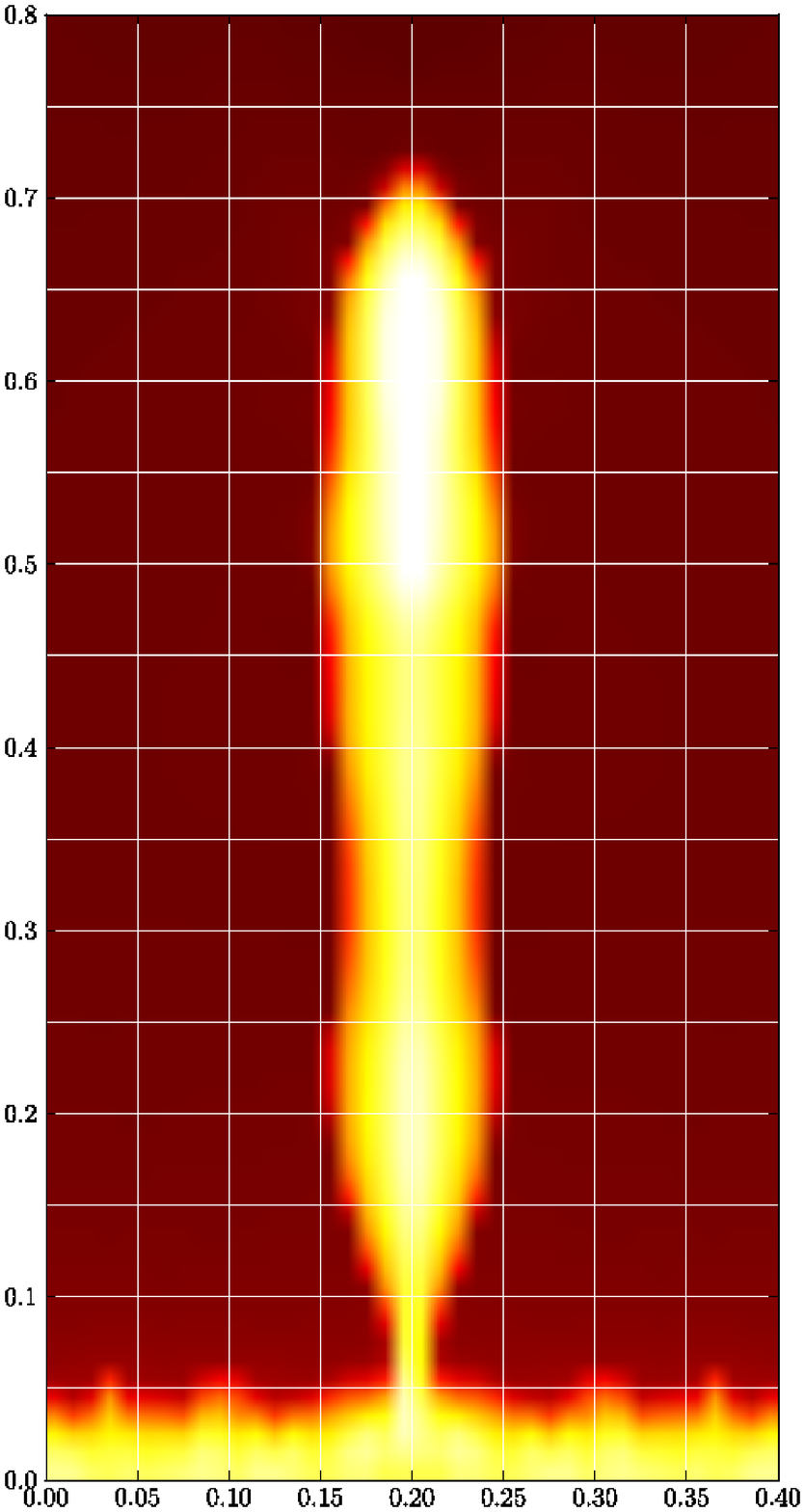}
\includegraphics[trim=1.8cm 3cm 1.5cm 3cm,clip,width=0.49\linewidth]{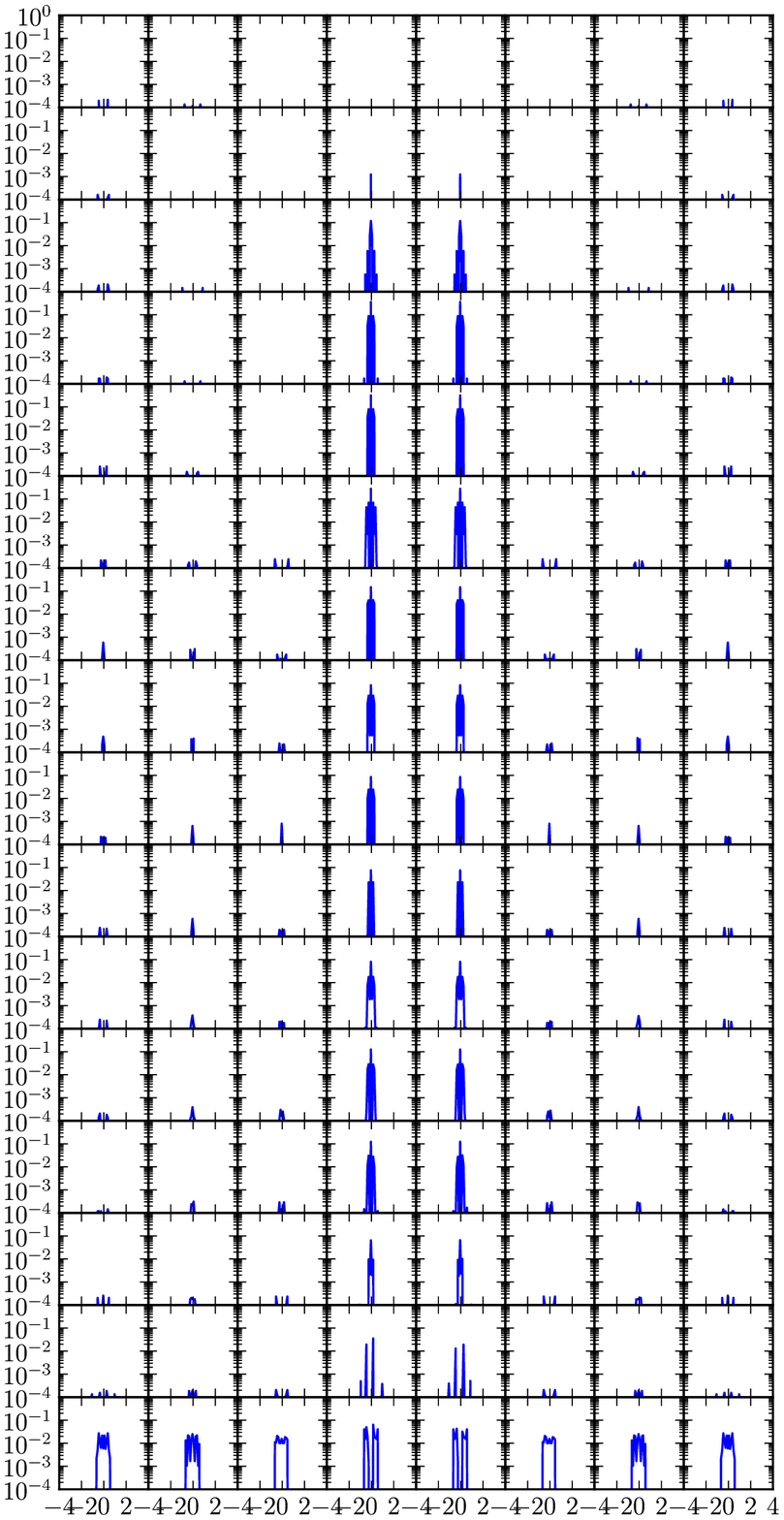}

\caption{\label{spectrum_middle} nH$_{clump}$/nH$_{cloud}$=5 simulation at t=444 ky. Same plots as Fig. \ref{spectrum_start}. The lateral shocks collided and cannot been identified anymore.}
\end{figure}

\begin{figure}

\centering
\includegraphics[trim=1.8cm 5.1cm 1.5cm 5.1cm,clip,width=0.49\linewidth]{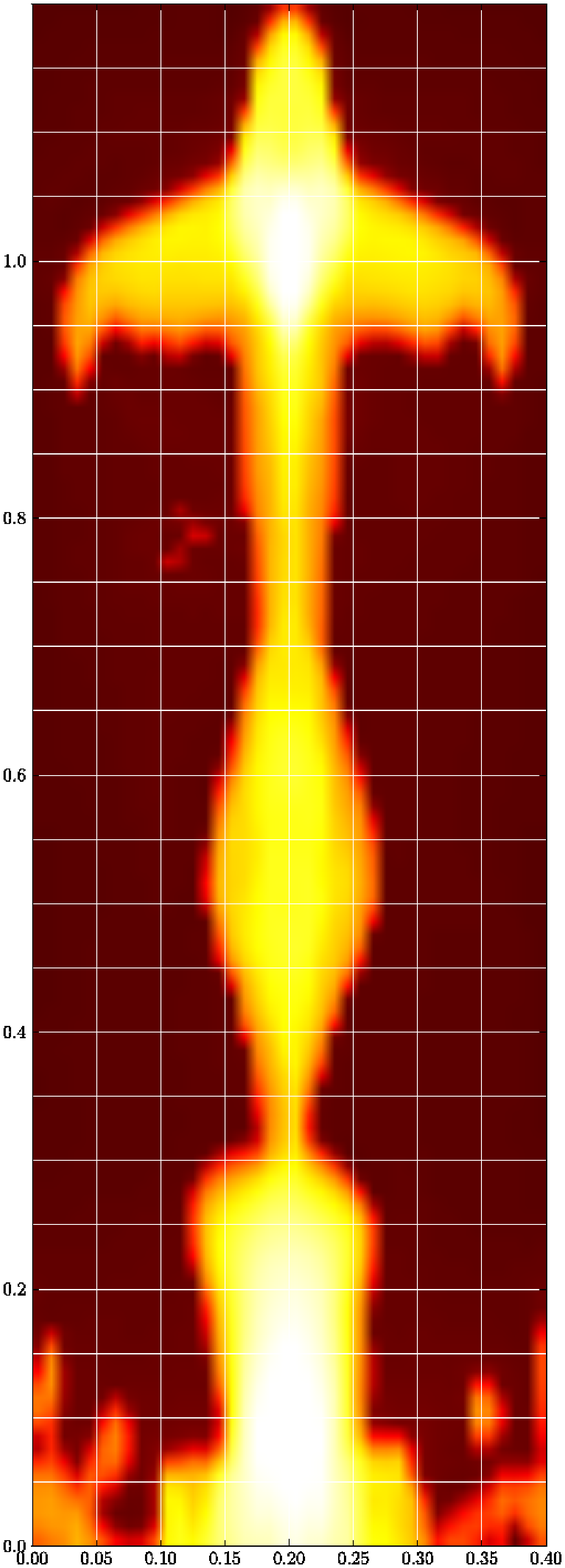}
\includegraphics[trim=1.8cm 4.5cm 1.5cm 4.5cm,clip,width=0.49\linewidth]{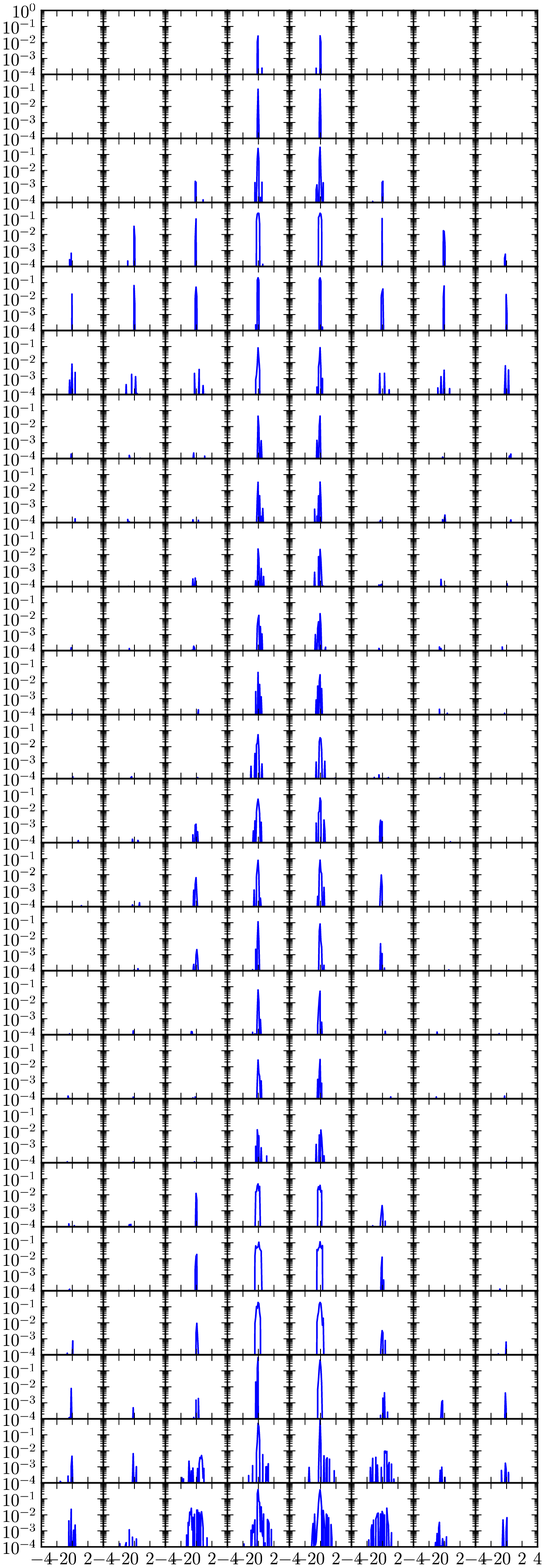}

\caption{\label{spectrum_end} nH$_{clump}$/nH$_{cloud}$=5 simulation at t=998 ky. Same plots as Fig. \ref{spectrum_start}. The line width is very small compared to the snapshots at t=222 ky.}
\end{figure}

The two sections above present how  pillars can be formed on a density
or interface  modulated region of  the molecular cloud.  These pillars
are  connected  to  the  cloud,  increasing  in  size  and  mass  with
time.  However these  variations can  not be  observed.  A potentially
observational signature to study is the structure of the line of sight
velocity.  We  use the  same  method  as \citet{Gritschneder2010}.  We
define a grid  of squares of 0.05x0.05 pc$^2$ along  the pillar and we
bin   the    line   of   sight   velocity   in    each   square   (see
Fig.       \ref{spectrum_start},       \ref{spectrum_middle}       and
\ref{spectrum_end}). We  use the simulation with the  densest clump to
study the line  of sight velocity structure but  the following results
are  generic. At  t=222  ky (Fig.  \ref{spectrum_start}), the  lateral
shocks can be clearly identified in the broad line spectrum. There are
two components,  a positive one coming  towards us and  a negative one
going away.  At this time the shell  is curved around a  hill which is
comparable to the situation we got in the interface-modulated scenario
after a short time.

At  t=444 ky  (Fig.  \ref{spectrum_middle}),  the lateral  shocks have
collided resulting in a small line width for the histograms. The whole
pillar is narrow and  evacuates the pressure through radiative cooling
and expansion to  reach the equilibrium defined for  the cold phase in
figure \ref{histo_cs03}.  There  is two dense part in  the pillar: the
head in  which the matter  of the clump  has been accumulated  and the
base at the point where the lateral shocks have closed. It will resist
the ionization  because it is dense  enough and matter  is starting to
accumulate at the base.

\begin{figure}

\centering
\includegraphics[trim=1.8cm 6.25cm 1.5cm 2cm,clip,width=0.49\linewidth]{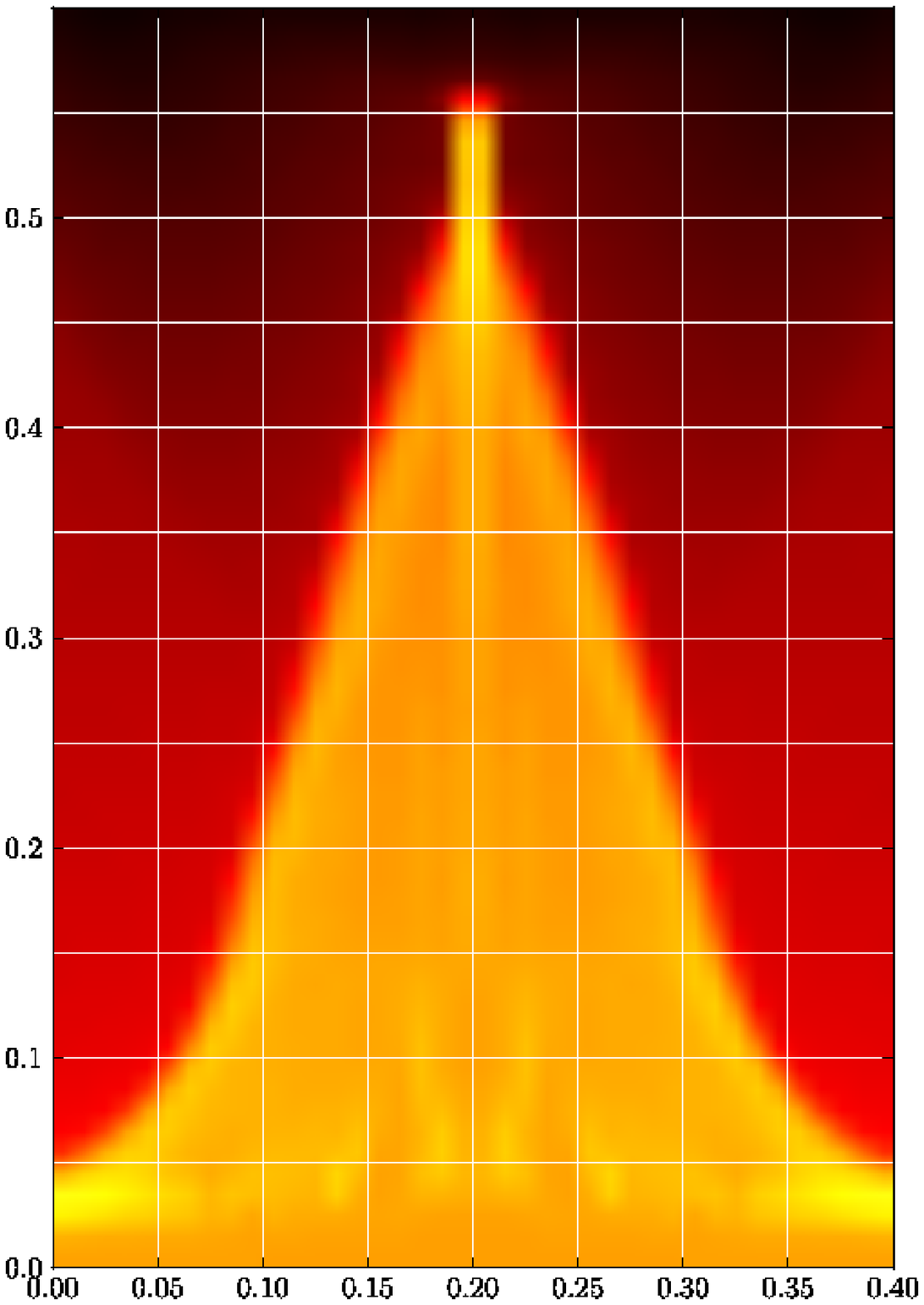}
\includegraphics[trim=1.8cm 2.1cm 1.5cm 2cm,clip,width=0.49\linewidth]{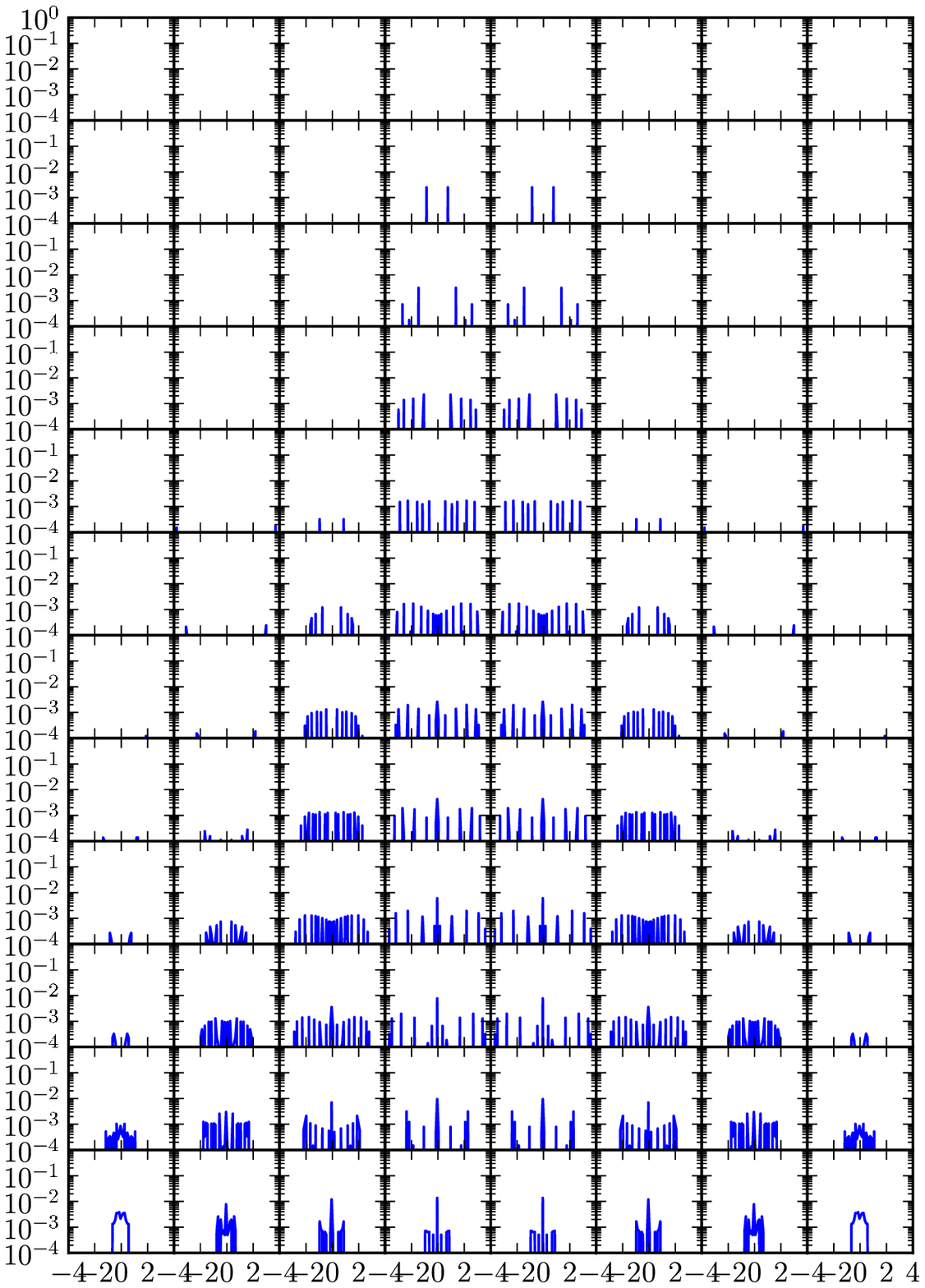}

\caption{\label{spectrum_c03_1} w/h=0.9 simulation at t=22.2 ky (first column in Fig. \ref{curve}). The shock just formed on the modulation, the line width and the double-wing structure is comparable to Fig. \ref{spectrum_start} }
\end{figure}

\begin{figure}

\centering
\includegraphics[trim=2cm 2.5cm 1.5cm 2cm,clip,width=0.49\linewidth]{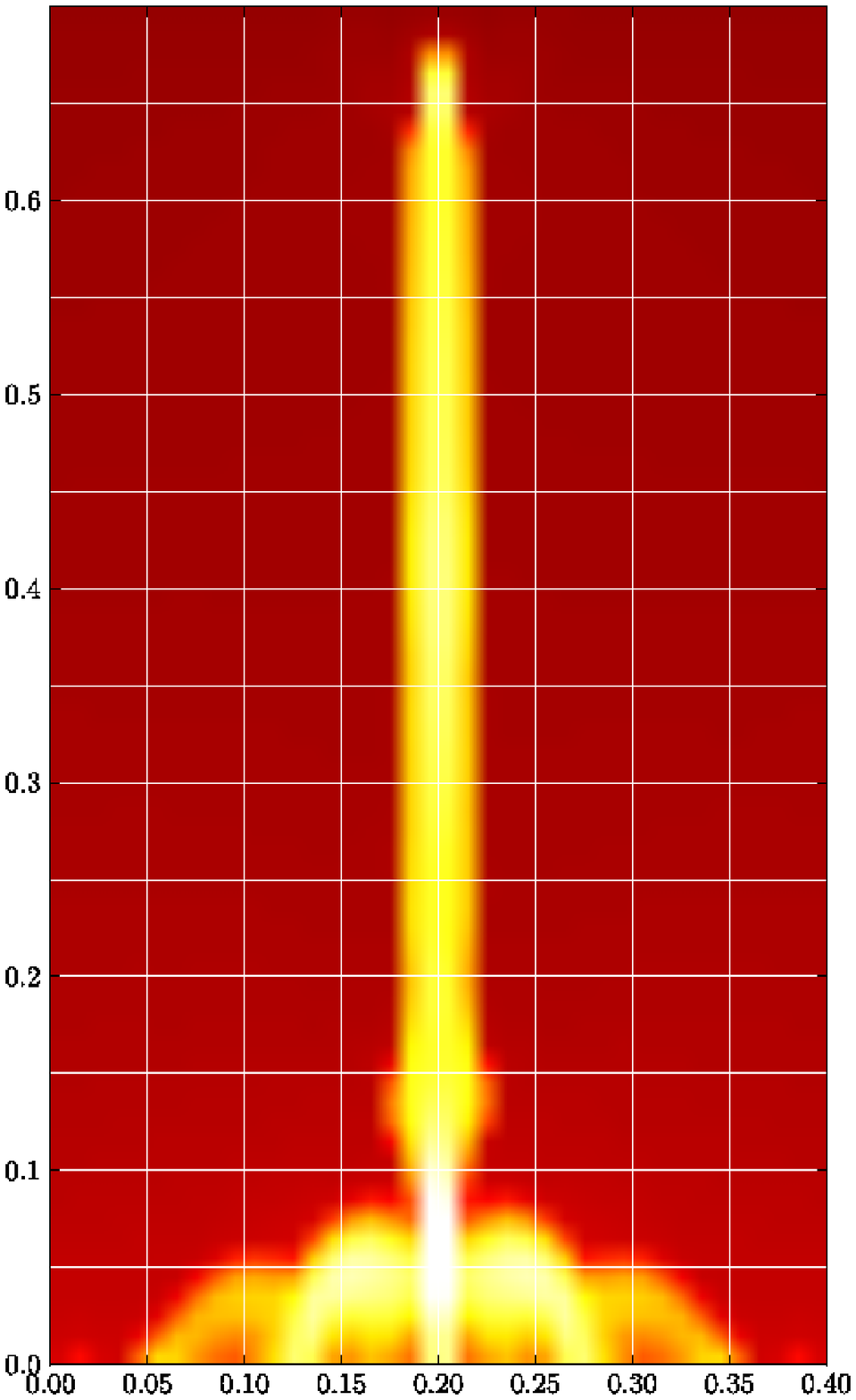}
\includegraphics[trim=1.8cm 2.5cm 1.5cm 2cm,clip,width=0.49\linewidth]{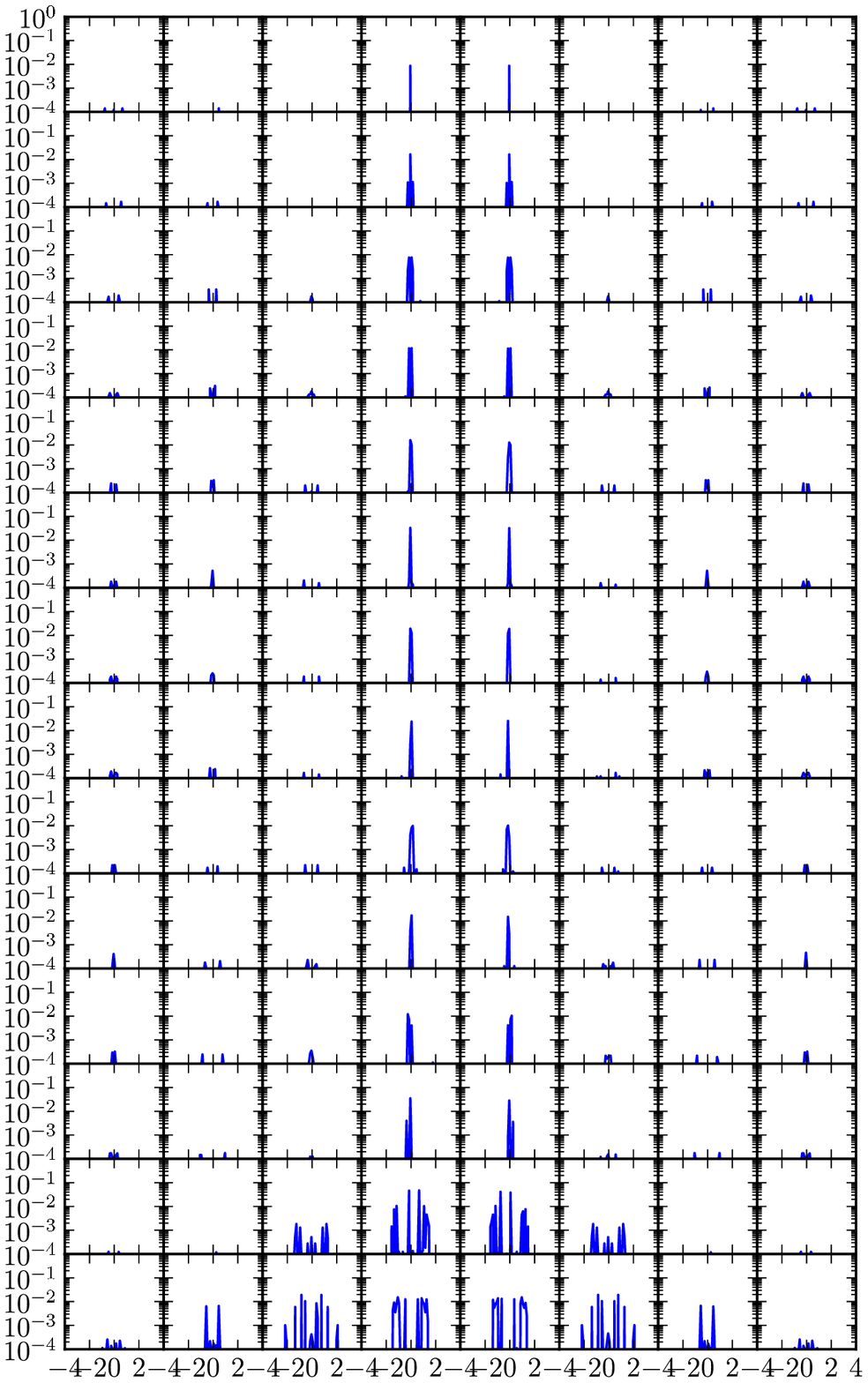}

\caption{\label{spectrum_c03_2} w/h=0.9 simulation at t=133.2 ky (third column in Fig. \ref{curve}). The shock collided lateraly and the line width is small as in Fig. \ref{spectrum_middle}}
\end{figure}

At t= 998 ky (Fig.  \ref{spectrum_end}), matter has accumulated at the
base  of the  pillar,  based on  the  process that  was identified  in
Fig. \ref{curve}.  The line of  sight velocity histograms have still a
small line width compared to Fig.\ref{spectrum_start}. This phenomenon
is  not specific  to  clumps  simulations, the  same  analysis on  the
interface   modulated   simulation    (aspect   ratio   of   0.9)   in
Fig. \ref{spectrum_c03_1} and  Fig. \ref{spectrum_c03_2} shows similar
results but  at an earlier  stage. The interface  modulated simulation
presents  double-wing velocity  spectra around  t=22.2 ky  whereas the
clumpy  simulation presents  these spectra  around t=222  ky  when the
shock  is curved  around  a  hill. The  double-wing  structure of  the
velocity spectra is therefore a  signature of the lateral shocks which
are going to  collide and a signature of the early  stage of a forming
pillar. \\

The  previous analysis  can be  applied to  the  different simulations
presented in this paper without noticable change to our conclusion. We
also  changed some  physical  conditions and  parameters.  We added  a
constant external gravitational  field, included self-gravity, doubled
the resolution without  global changes in the morphology  of the final
structures. We also  changed the flux to an  higher value (from 10$^9$
to  5$\times$10$^9$  photons/s/cm$^2$) and  the  structures  are similar  but
evolve  faster as  seen by  \citet{Gritschneder2009}. The  shell speed
$c_{shell}$     is    proportional     to     $F_\gamma^{1/4}$    (see
Eq. \ref{hii_param}  and Eq. \ref{shell_param})  therefore multiplying
the  flux by  five increases  the shell  speed by  50\% and  the whole
simulation evolves 50\% faster.  In all these situations, pillars form
by  the  lateral  collision  of  a  curved  shocked  surface  and  the
double-wing spectrum is always visible before the shocks collide.

%
%

\section{Conlusion and discussion}

We have presented a new scenario for the formation of structures at the
edge of HII regions and shown that:
\begin{enumerate}[i]
\item A curved shock ahead of an ionization front can lead to a pillar
if it is curved enough to collide lateraly on itself.
\item This pocess is very efficient to form stable growing pillar, the
narrower  the initial  structure, the  more curved  the front  and the
longer the pillar.
\item Lateral  gas flows  can result in  a density enhancement  at the
base of the pillars of 1-2 orders of magnitude compared to the collect
and collapse scenario.
\item  When the  shock  is first  formed  flat, it  can  be curved  by
enough-dense  clumps   and  leads  to  pillars.   On  isolated  clumps
(Radiation Driven  Implosion) , the  shock is naturally curved  by the
form of the clump but the  resulting structure has a constant size and
mass.
\item The double-wing line spectra  of the line-of-sight velocity is a
signature  of the  lateral  collision of  the  shock, and  hence is  a
signature of the  early stage of the formation of a  pillar. It can be
used as an observational signature for this new scenario.
\end{enumerate}

Various aspects  of shocks orientation  have been considered  by other
authors.    Oblique   shocks   have   previously   been   studied   by
\citet{Chevalier1975}       and       cylindrical      shocks       by
\citet{KimuraToshiya1990}. These effects have been further explored in
the context of  2D turbulent simulations \citep[see][]{Elmegreen1995}.
However  these studies  focus on  the effects  of curvature  on clumps
enhancement.   In this  work, we  have shown  that shocks  need  to be
enough curved to collide on themselves  in order to form a pillar like
structure.

At  the  edge of  HII  regions,  the  structure of  the  line-of-sight
velocity  can  be  investigated  using  radiotelescopes  and  suitable
molecules tracing  the dynamic, to  detect the double-wing  spectra on
gas hills and  thus nascent pillars. If a  gas overdensity is detected
at the top of the hill (e.g. using Herschel data), the shock curvature
could be attributed to the presence of an initial clump, if not to the
curvature of an initial interface.\\

In this study, curved shocks  have been generated using either density
or  interface modulations  on  a pc  scale  and in  a very  simplified
situation. However, we expect that the physical processes discussed in
this paper can also be  applied to more realistic situations where the
ionization  front interact  with a  turbulent interstellar  medium. In
this case the density fluctuations are much more complex, but the same
mechanism  should  be at  work  to  describe  pillar formation.  These
turbulent situations will  be studied in a forthcoming  paper in which
the interplay  between shock curvature and turbulence  will be studied
to  see  its  impact on  star  formation  rates  at  the edge  of  HII
regions.\\

%
%


%
%

\bibliographystyle{aa}
\bibliography     {bibliography.bib}


\end {document}